%% file: main.tex
\documentclass[12pt,table]{elsarticle}

\usepackage[utf8]{inputenc}
\usepackage{amsmath}
\usepackage{amssymb}
\usepackage{graphicx}
\usepackage{todonotes}
\usepackage{subcaption}
\usepackage{algorithm}
\usepackage{algorithmic}
\DeclareMathOperator*{\argmin}{argmin}
\usepackage{multirow}
\usepackage{xcolor}
\usepackage{hyperref}
\usepackage{enumitem}
\usepackage{url}
\usepackage{afterpage}
\usepackage{lscape}
\usepackage{amssymb}
\usepackage{rotating}
\usepackage{booktabs}
\usepackage{soul}
\usepackage{xparse}

\NewDocumentCommand{\rot}{O{45} O{1em} m}{\makebox[#2][l]{\rotatebox{#1}{#3}}}%

\definecolor{lightblue}{rgb}{.60,.7,1}
\sethlcolor{lightblue}

\graphicspath{{kuvat/}}

\makeatletter
\AtBeginDocument{\let\hl\@firstofone}
\makeatother

\begin{document}

\input{Elsevier_cover.tex}

\begin{frontmatter}

\title{Edge computing server placement with\\capacitated location allocation}

\author[mat]{Tero Lähderanta\corref{cor1}}
\author[ubi]{Teemu Leppänen\fnref{label2}}
\author[mat,luke]{Leena Ruha\fnref{label2}}
\author[ubi]{Lauri Lovén\fnref{label2}}
\author[cwc]{Erkki Harjula}
\author[cwc]{Mika Ylianttila}
\author[ubi]{Jukka Riekki}
\author[mat]{Mikko J. Sillanpää}
\address[mat]{Research Unit of Mathematical Sciences, University of Oulu, Finland}
\address[ubi]{Center for Ubiquitous Computing,  University of Oulu, Finland}
\address[cwc]{Centre for Wireless Communications, University of Oulu, Finland}
\address[luke]{Natural Resources Institute Finland, Oulu, Finland}
\fntext[label2]{These authors contributed equally}
\cortext[cor1]{corresponding author: tero.lahderanta@oulu.fi \\ P.O.Box 8000 FI-90014 University of Oulu}
\begin{abstract}
The deployment of edge computing infrastructure requires a careful placement of the edge servers, with an aim to improve application latencies and reduce data transfer load in opportunistic Internet of Things systems. In the edge server placement, it is important to consider computing capacity, available deployment budget, and hardware requirements for the edge servers and the underlying backbone network topology.
In this paper, we thoroughly survey the existing literature in edge server placement, identify gaps and present an extensive set of parameters to be considered. We then develop a novel algorithm, called PACK, for server placement as a capacitated location-allocation problem. PACK minimizes the distances between servers and their associated access points, while taking into account capacity constraints for load balancing and enabling workload sharing between servers. Moreover, PACK considers practical issues such as prioritized locations and reliability. We evaluate the algorithm in two distinct scenarios: one with high capacity servers for edge computing in general, and one with low capacity servers for Fog computing. Evaluations are performed with a data set collected in a real-world network, consisting of both dense and sparse deployments of access points across a city area. The resulting algorithm and related tools are publicly available as open source software.
\end{abstract}


%
%
%
%

\begin{keyword}

multi-access edge computing, 
facility location,
clustering,
k-medoid

\end{keyword}

\end{frontmatter}


\input{1_introduction.tex}

\input{2_related_work.tex}

\input{3_method.tex}

\input{4_evaluation.tex}

{
\afterpage{
\begin{figure*}[ht!]
\centering
\captionsetup[subfigure]{labelformat=empty}
\begin{subfigure}[t]{.45\textwidth}
  \centering
  \includegraphics[width=1\linewidth]{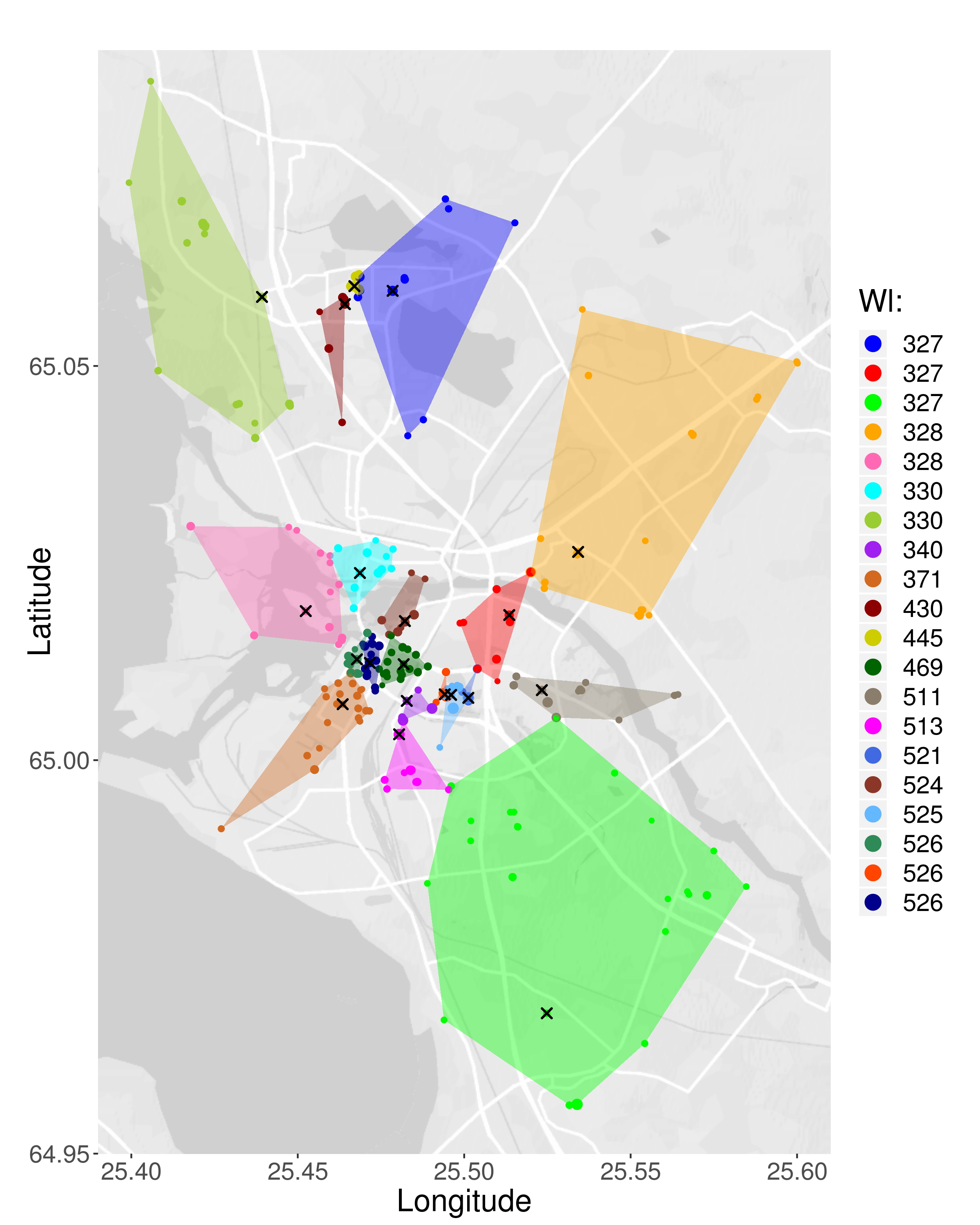}
 \caption{\textbf{\textit{M1} Upper and lower capacity constraints.} \\ \hl{Spatially large server regions, but well balanced workloads.}}
\end{subfigure}
\begin{subfigure}[t]{.45\textwidth}
  \centering
  \includegraphics[width=1\linewidth]{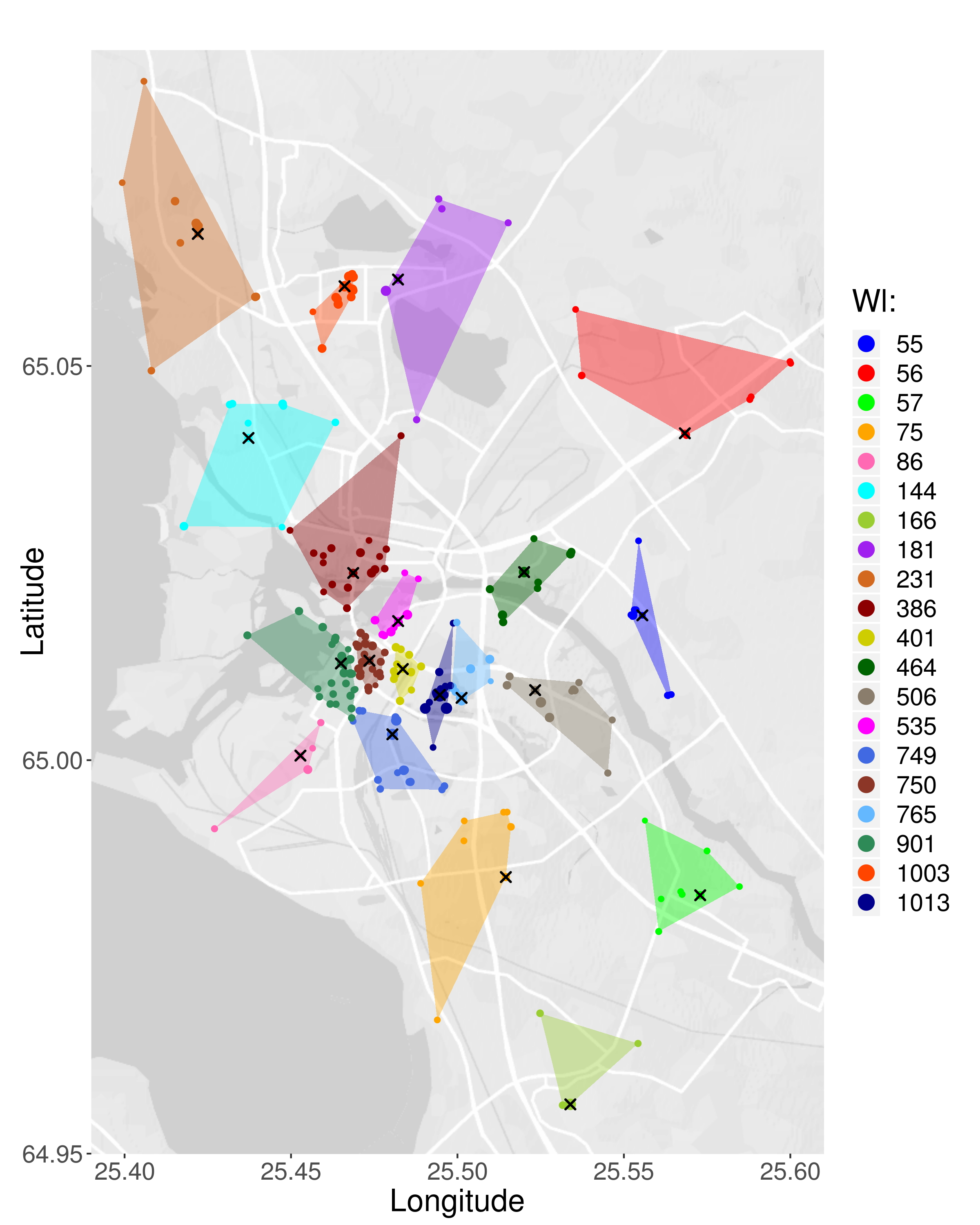}
 \caption{\textbf{\textit{M2} No capacity constraints.} \\ \hl{Very uneven distribution of workload, but spatially compact server areas.}}
\end{subfigure}
\begin{subfigure}[t]{.45\textwidth}
  \centering
 \includegraphics[width=1\linewidth]{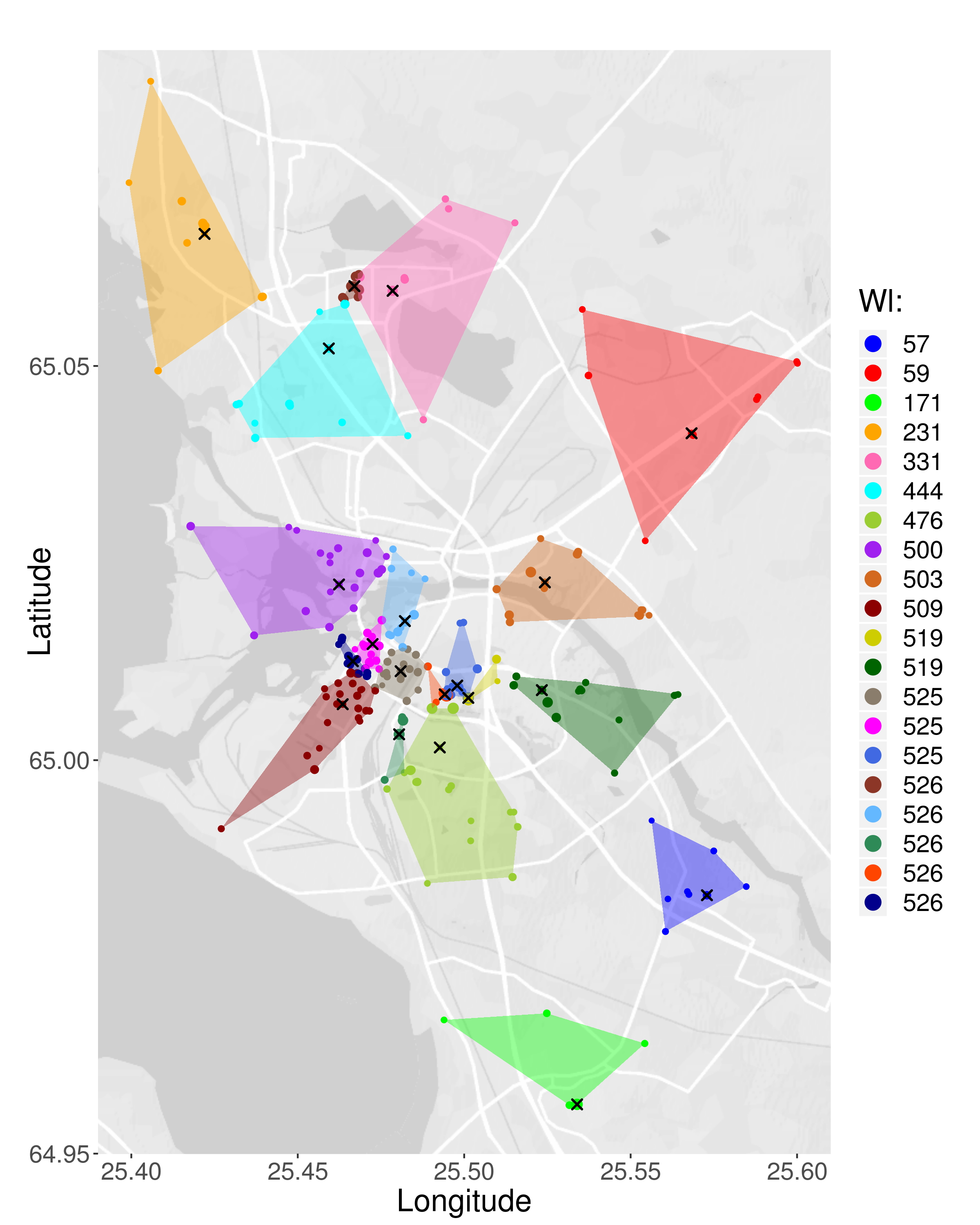}
  \caption{\textbf{\textit{M3} Only upper capacity constraint.} \\ \hl{Underused capacity in many servers, but no exceedingly high workloads.}}
\end{subfigure}

\caption{MEC server placement scenarios. Crosses denote the server locations. Wl: workloads of the servers. The color shades visualize the server regions. \hl{Quantitative measures of the scenarios can be seen in Table} \ref{QoS_table_MEC}.}
\label{mec_scenarios}
 \end{figure*}
\clearpage
}

\afterpage{
\begin{figure*}[ht!]
\centering
\captionsetup[subfigure]{labelformat=empty}
\begin{subfigure}[t]{.45\textwidth}
  \centering
  \includegraphics[width=1\linewidth]{kuvat/mec-k20.jpg}
 \caption{\textbf{\textit{M1} Upper and lower capacity constraints.} \\ \hl{Scenario with no priority APs.} }
\end{subfigure}
\begin{subfigure}[t]{.45\textwidth}
  \centering
 \includegraphics[width=1\linewidth]{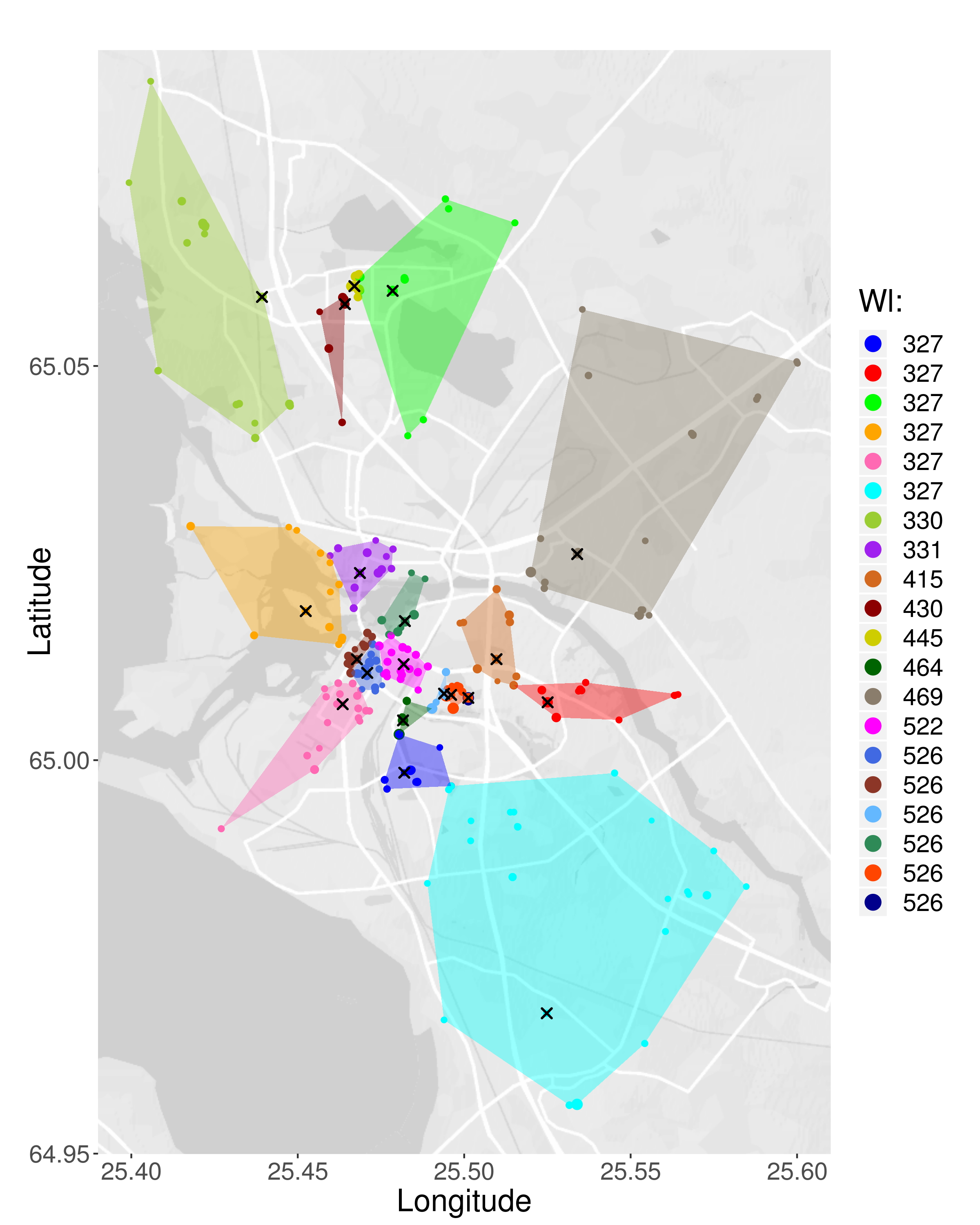}
  \caption{\textbf{\textit{M4} Fractional membership.} \\ \hl{Similar to \textit{M1}, but allows the use of shared workload.}}
\end{subfigure}
\begin{subfigure}[t]{.45\textwidth}
  \centering
  \includegraphics[width=1\linewidth]{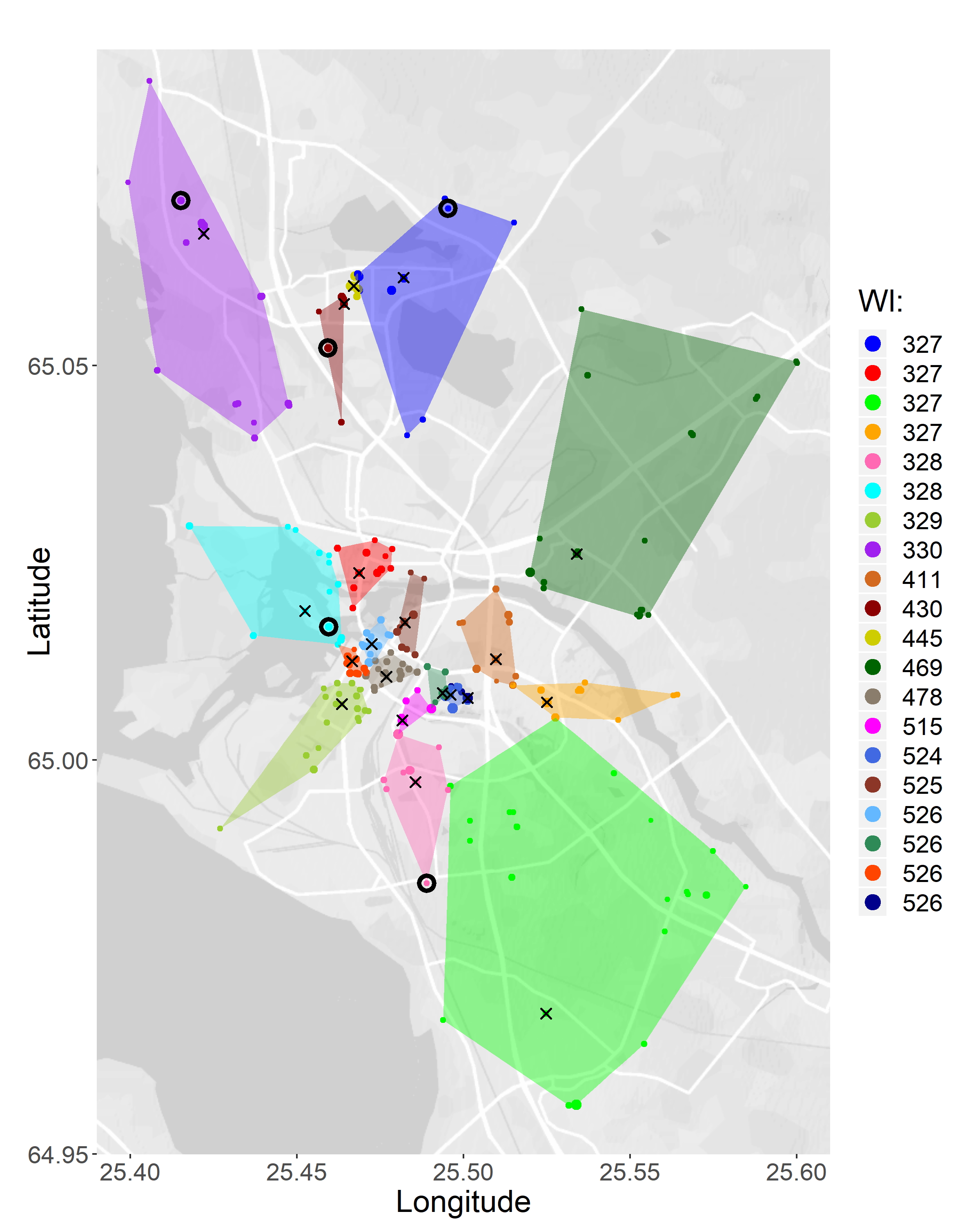}
 \caption{\textbf{\textit{M5} Location priority scenario.} \\ \hl{Edge servers are closer to the priority APs.}}
\end{subfigure}
\begin{subfigure}[t]{.45\textwidth}
  \centering
  \includegraphics[width=1\linewidth]{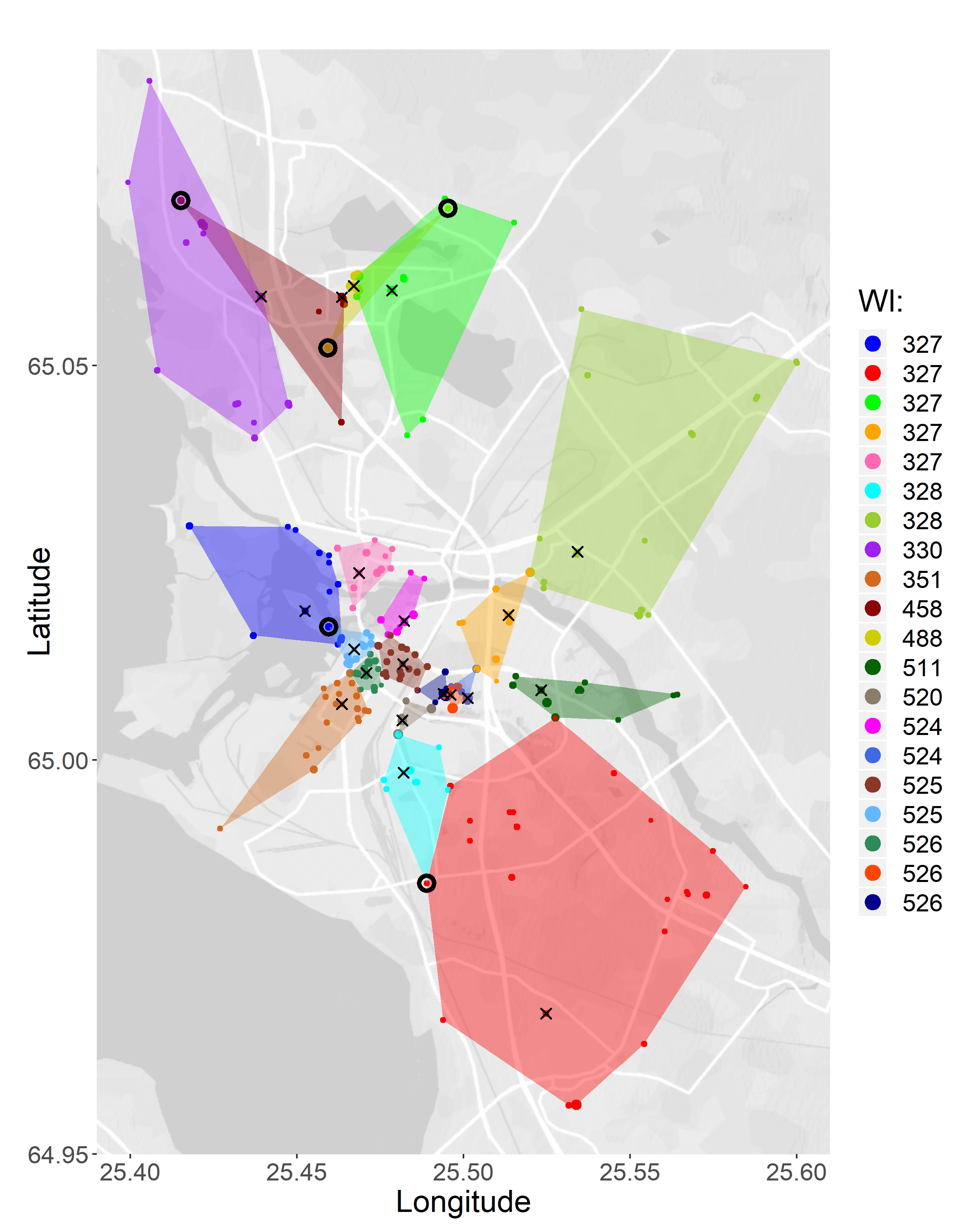}
 \caption{\textbf{\textit{M6} Reliability scenario.} \\ \hl{APs are assigned to two servers, thus producing overlapping clusters}}
\end{subfigure}
\caption{Comparing placement scenario with location priority to the \textit{M1} scenario. Crosses denote the server locations and circles denote the APs with priority. Wl: workloads of the servers. The color shades visualize the server regions. \hl{Quantitative measures of the scenarios can be seen in Table} \ref{QoS_table_MEC}.}
\label{mec_priority}
 \end{figure*}
\clearpage
}
\afterpage{
 \begin{figure*}[ht!]
\centering
\captionsetup[subfigure]{labelformat=empty}
\begin{subfigure}[t]{.4\textwidth}
  \centering
 \includegraphics[width=1\linewidth]{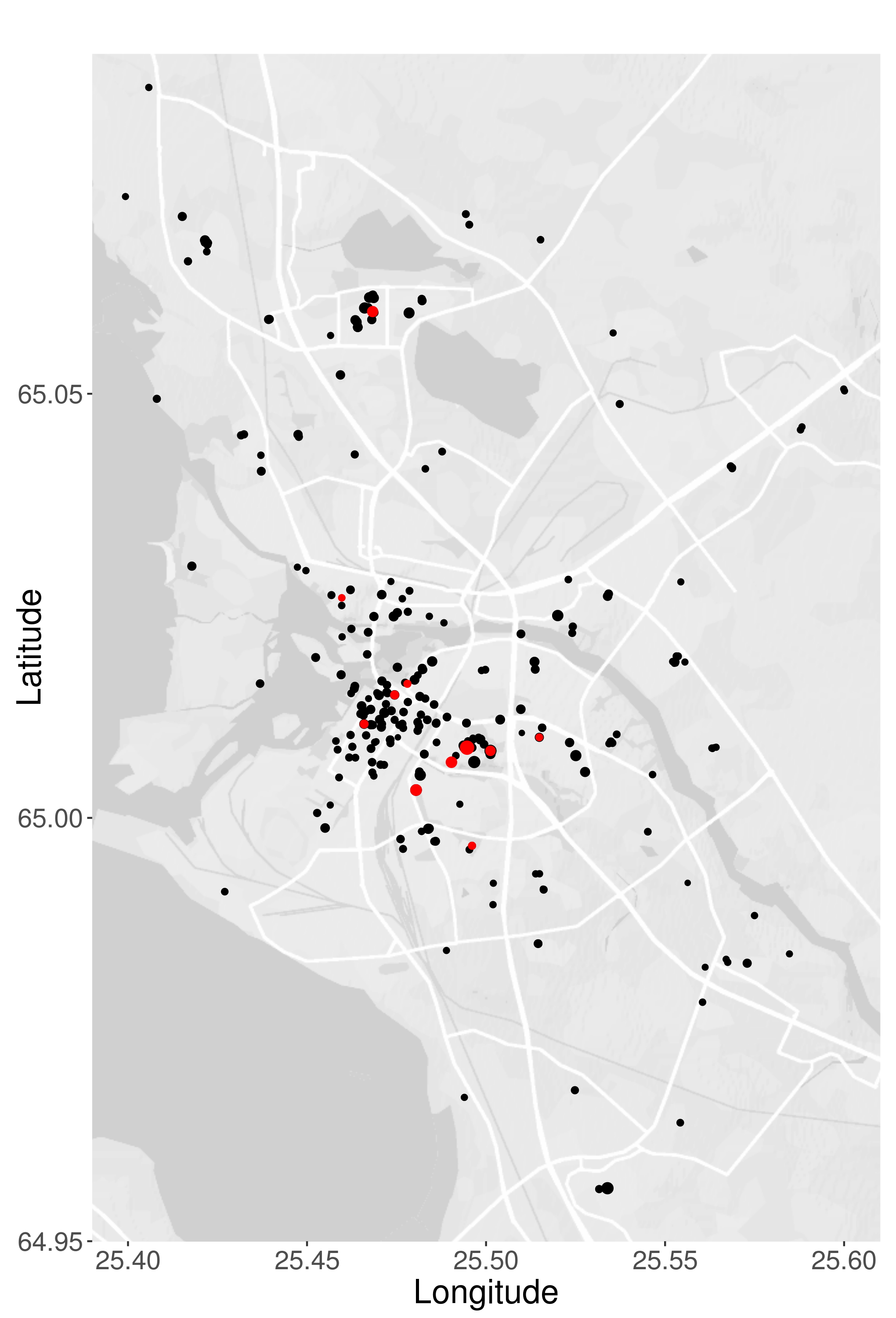}
  \caption{\textbf{\textit{M4} MEC with fractional membership.} \hl{Less shared workload with fewer servers.}}
\end{subfigure}
\begin{subfigure}[t]{.4\textwidth}
  \centering
 \includegraphics[width=1\linewidth]{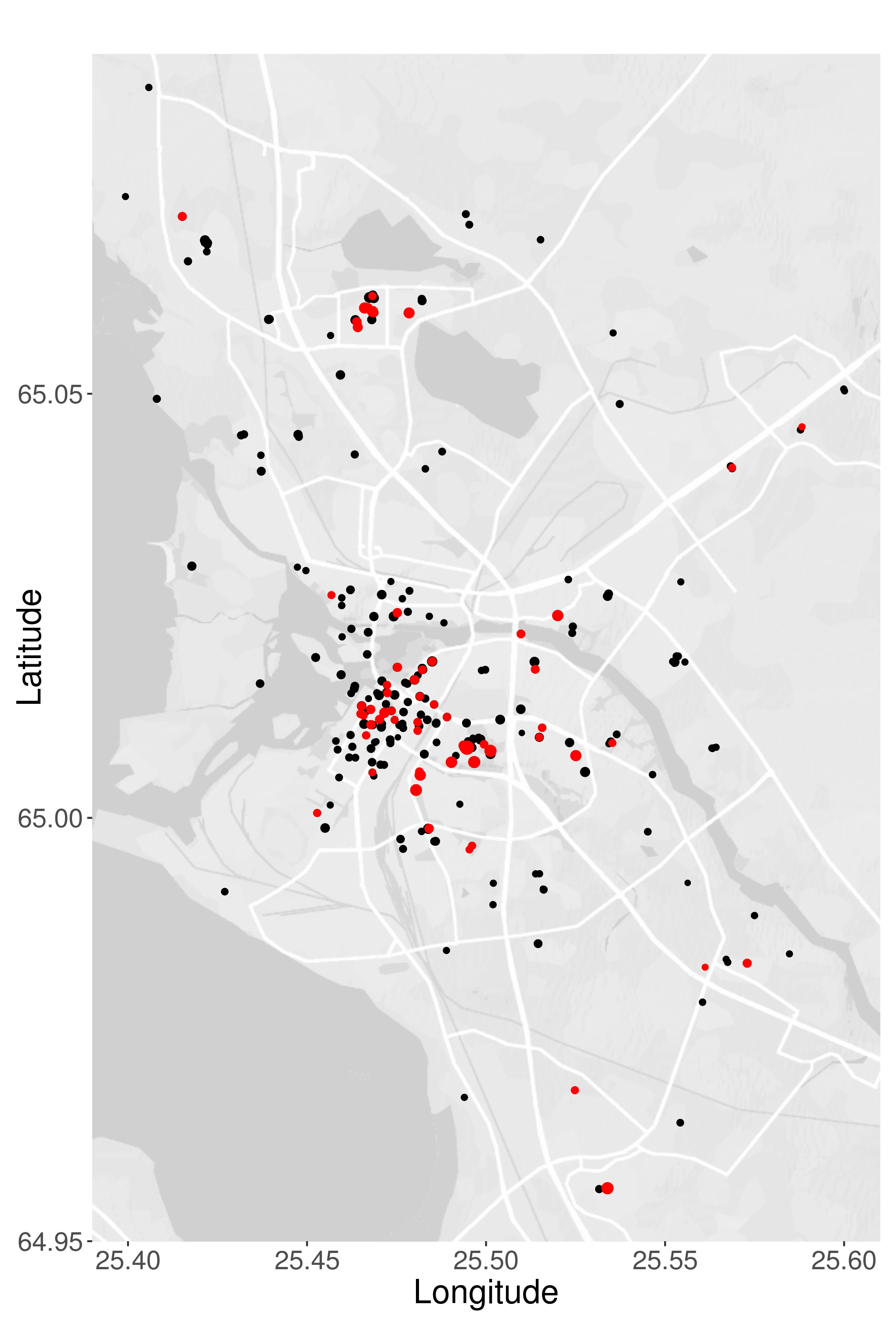}
  \caption{\textbf{\textit{F1} Fog with upper and lower capacity constraints.} \hl{More shared workload all across the spatial region.}}
\end{subfigure}
\begin{subfigure}[t]{.4\textwidth}
  \centering
 \includegraphics[width=1\linewidth]{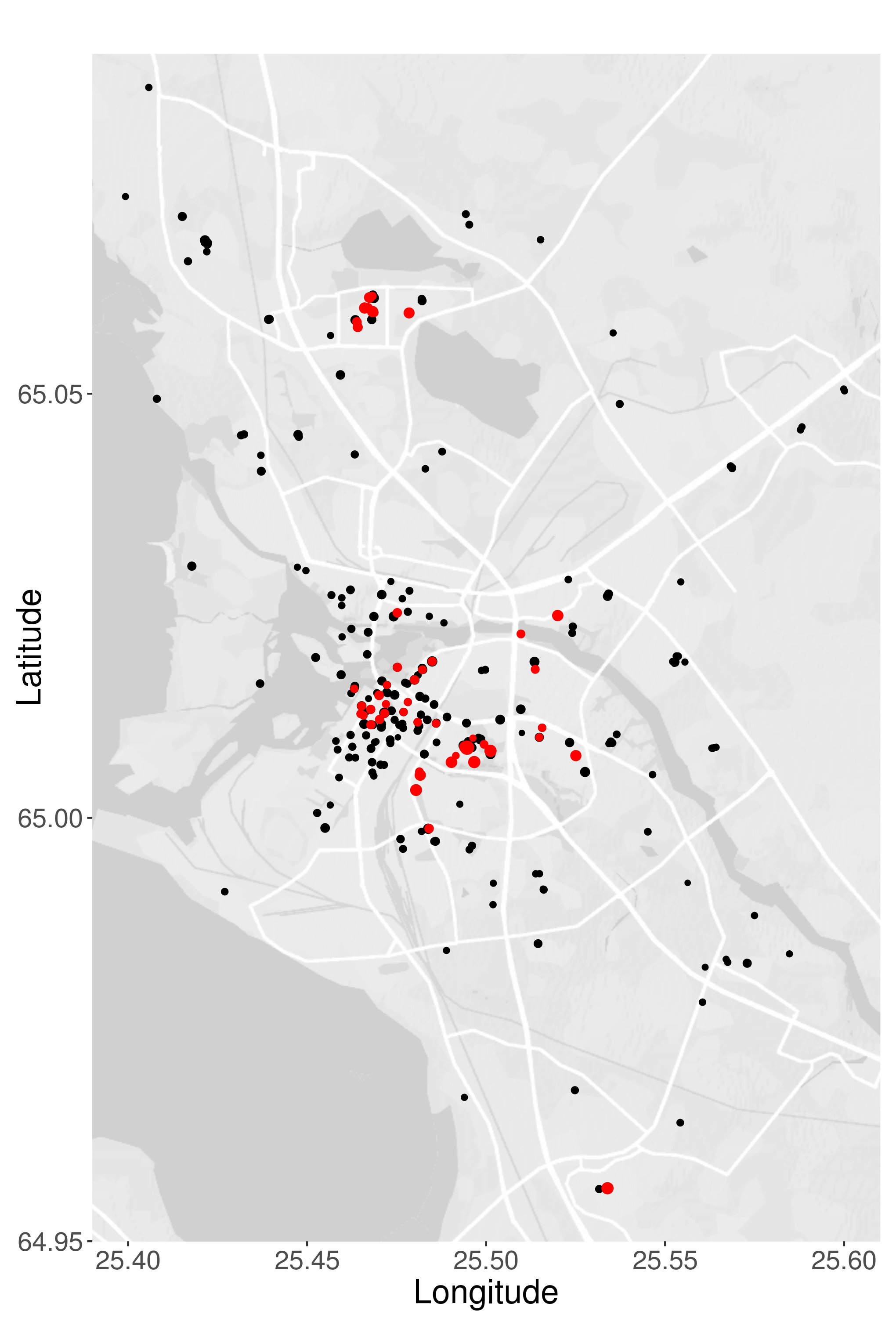}
  \caption{\textbf{\textit{F3} Fog with only upper capacity constraints.} \hl{More shared workload in the dense regions and less in the sparse regions.}}
\end{subfigure}
\caption{APs with shared workloads in fractional membership scenarios. Black: the AP is assigned to exactly one server. Red: the workload of the AP is shared across servers.}
\label{split_workload}
 \end{figure*}
\clearpage
}
\afterpage{
\begin{figure*}[ht!]
\centering
\captionsetup[subfigure]{labelformat=empty}
\begin{subfigure}[t]{.45\textwidth}
  \centering
 \includegraphics[width=1\linewidth]{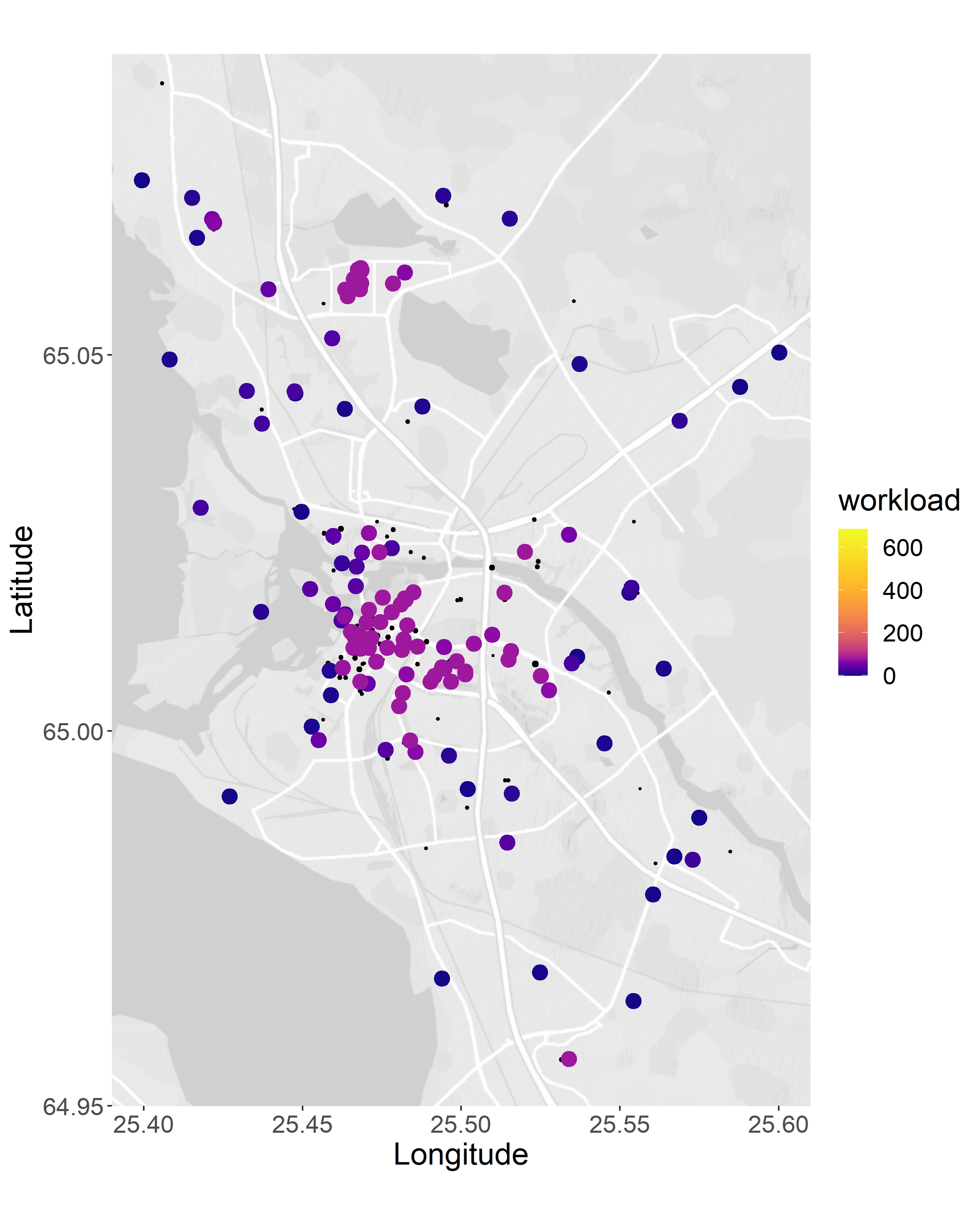}
  \caption{\textbf{\textit{F1} Upper and lower capacity constraints.} \\ \hl{Spatially large server regions, but well balanced workload between servers.}}
\end{subfigure}
\begin{subfigure}[t]{.45\textwidth}
  \centering
 \includegraphics[width=1\linewidth]{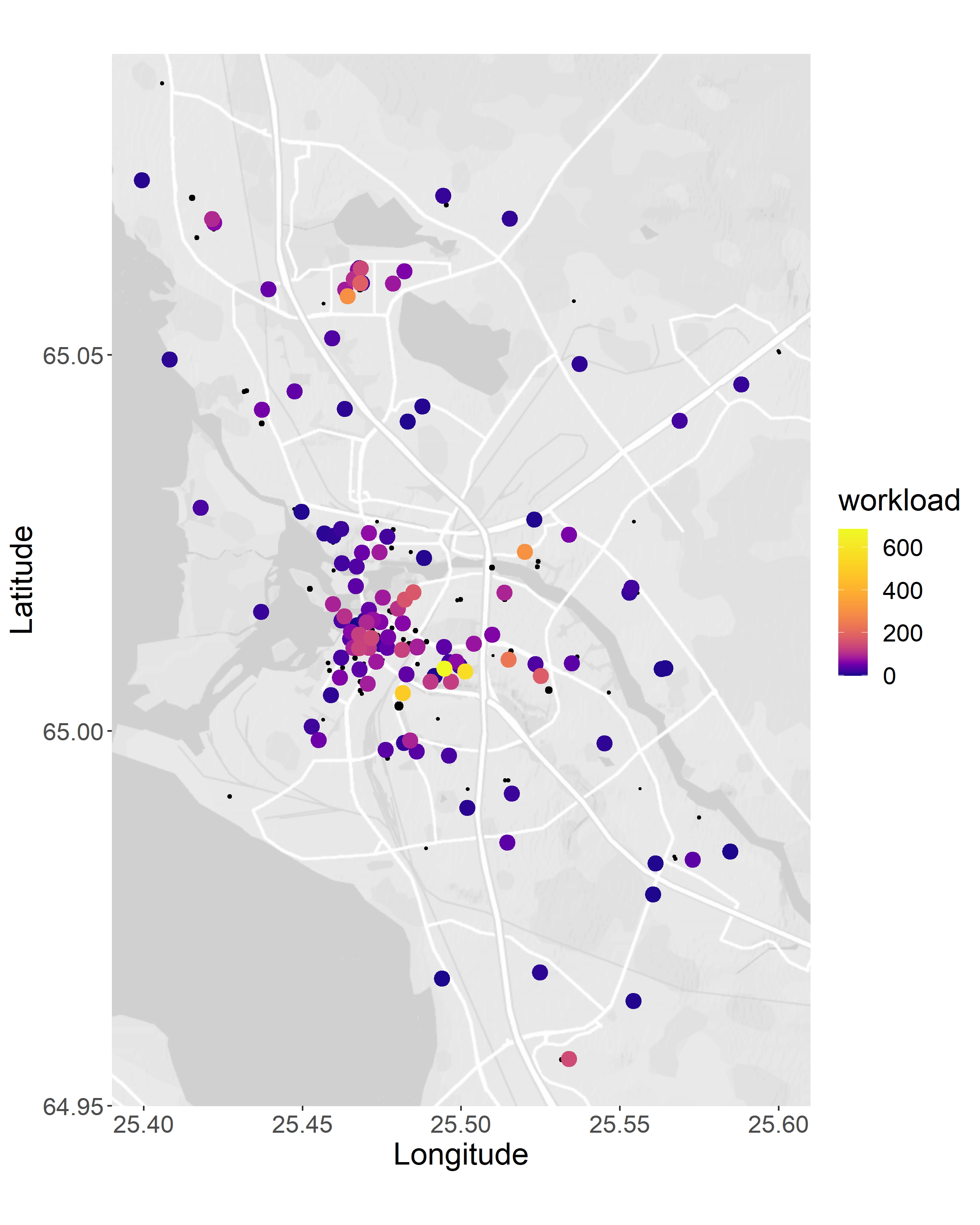}
  \caption{\textbf{\textit{F2} No capacity constraints with fractional membership.} \\ \hl{Uneven distribution of workload between servers, but overall better proximity.}}
\end{subfigure}
\begin{subfigure}[t]{.45\textwidth}
  \centering
 \includegraphics[width=1\linewidth]{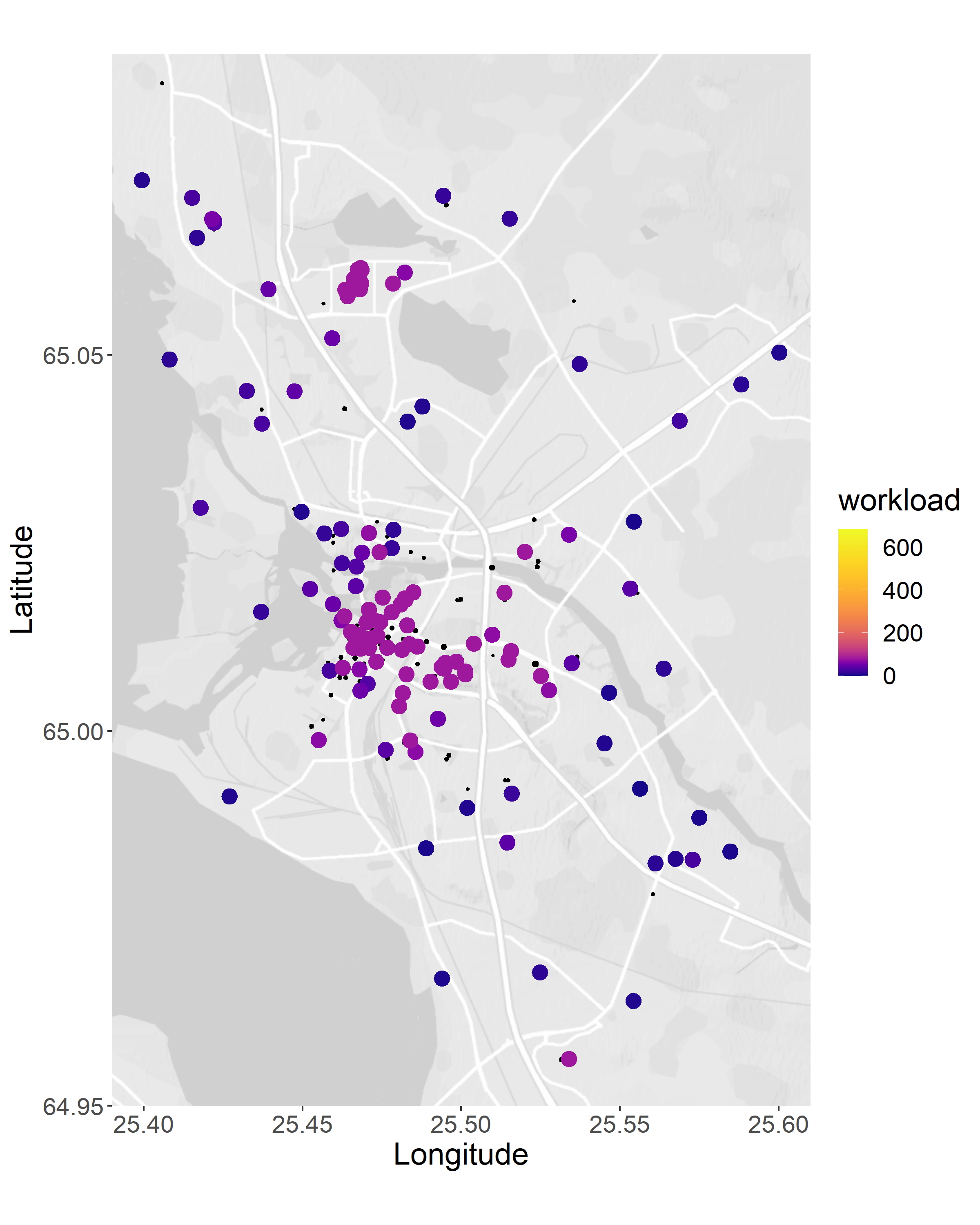}
  \caption{\textbf{\textit{F3} Only upper capacity constraint.} \\ \hl{Few servers with close to zero workload, but otherwise well balanced workload.}}
\end{subfigure}
\caption{Optimal placement of 150 Fog servers. Black dots indicate the APs and colored dots represent the servers, the  shade of color indicating the workload (Wl). \hl{Quantitative measures of the scenarios can be seen in Table} \ref{tab:QoS_table_Fog}.}
\label{Fog_clustering}
 \end{figure*}
\clearpage
}
\afterpage{
\begin{figure*}[ht!]
\centering
\captionsetup[subfigure]{labelformat=empty}
\begin{subfigure}[t]{.32\textwidth}
  \centering
 \includegraphics[width=1\linewidth]{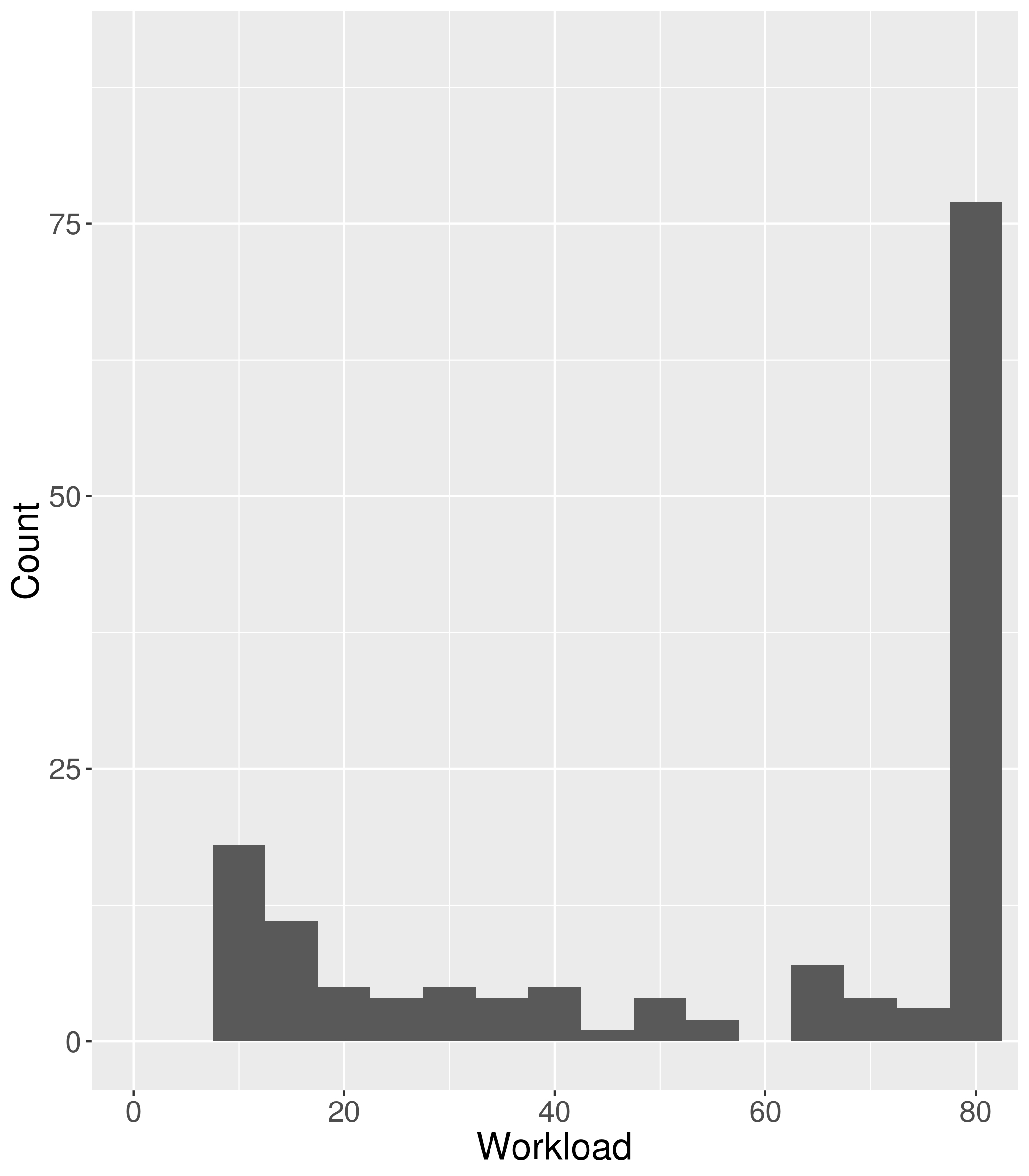}
  \caption{\textbf{\textit{F1} Upper and lower capacity constraints.}}
\end{subfigure}
\begin{subfigure}[t]{.32\textwidth}
  \centering
 \includegraphics[width=1\linewidth]{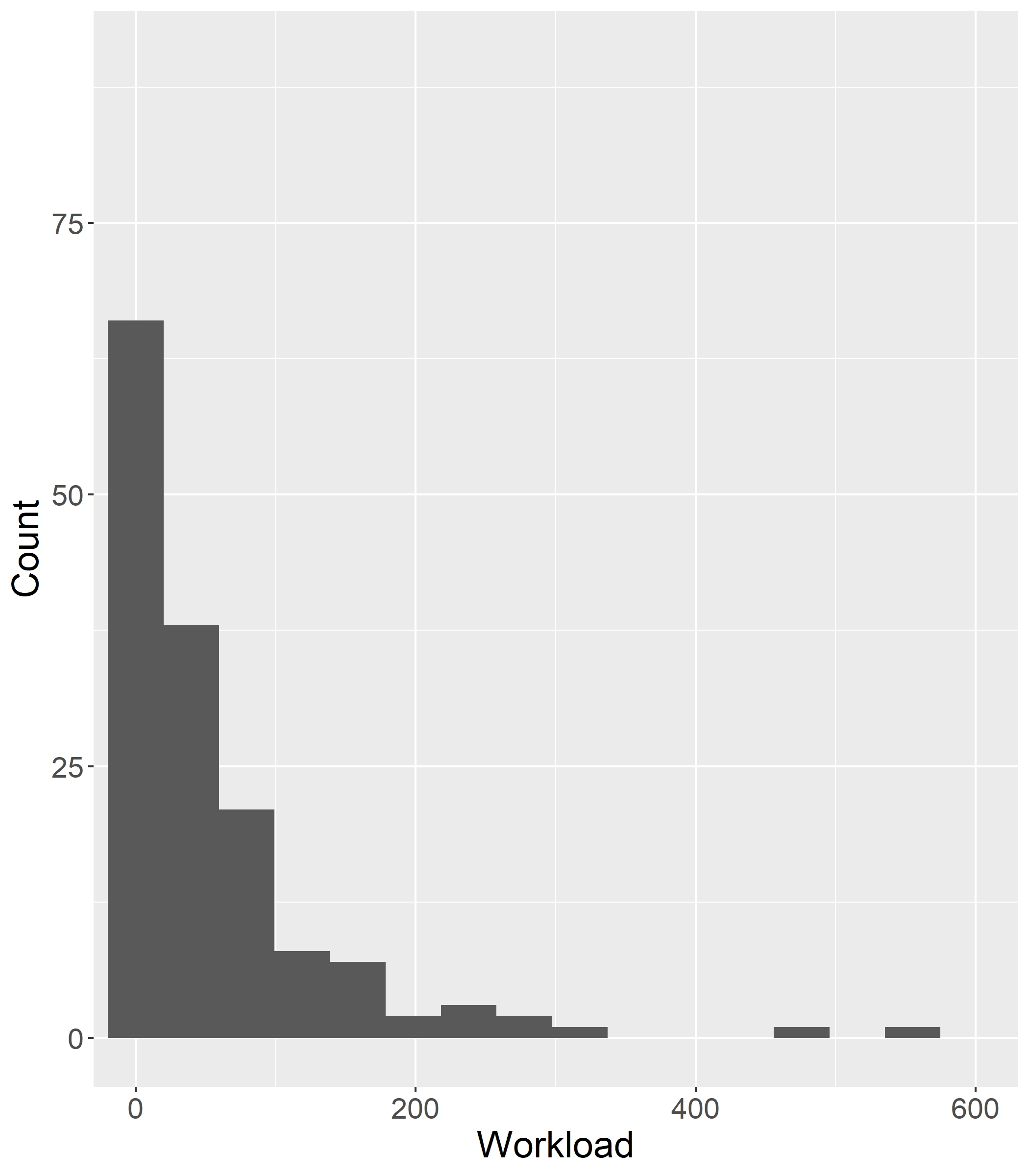}
  \caption{\textbf{\textit{F2} No capacity constraints with fractional membership.}}
\end{subfigure}
\begin{subfigure}[t]{.32\textwidth}
  \centering
 \includegraphics[width=1\linewidth]{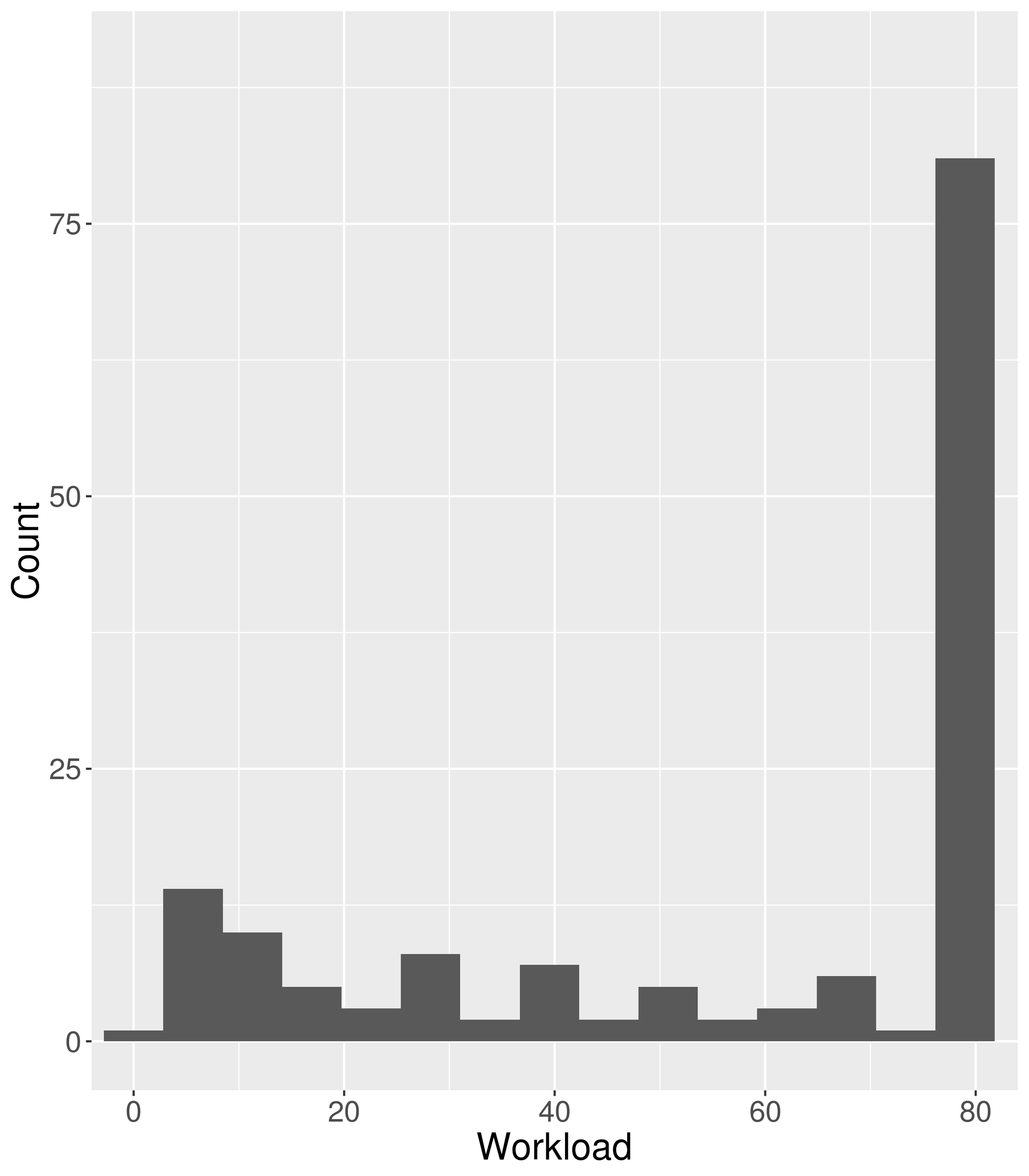}
  \caption{\textbf{\textit{F3} Only upper capacity constraint.}}
\end{subfigure}
\caption{Histograms of Fog server workloads in each scenario. \hl{The histograms show two peaks of workload in scenarios \textit{F1} and \textit{F3}, while workload in scenario \textit{F2} has only one peak.} Note different vertical and horizontal scale in \textit{F2}.}
\label{Fog_hist}
 \end{figure*}
\clearpage
}
}
\newpage 

\input{5_discussion.tex}

\input{6_conclusion.tex}

\input{7_acknowledgments.tex}

\bibliographystyle{elsarticle-num}
\bibliography{refs}{}

\input{appendix_tight.tex}

\end{document}

%% file: Elsevier_cover.tex
\pagenumbering{gobble}
\clearpage
\thispagestyle{empty}

\noindent
This is the accepted version of the paper "Edge server placement with capacitated location allocation" by Lähderanta et al. (2021).
The published version can be found in Journal of Parallel and Distributed Computing vol 153, July 2021, Pages 130-149. \url{https://doi.org/10.1016/j.jpdc.2021.03.007} \\ 

\vspace{5mm}
\noindent
© 2021. This manuscript version is made available under the CC-BY-NC-ND 4.0 license \url{http://creativecommons.org/licenses/by-nc-nd/4.0/}

\clearpage
\pagenumbering{arabic}

%% file: 1_introduction.tex


\section{Introduction}

Edge computing has emerged as a scalable and flexible solution to address the challenges introduced by cloud-centric Internet of Things (IoT) systems \cite{Yousefpour2019}. Cloud is physically and logically far away from the things, and requires massive upstream data uploads, consuming core network resources and introducing latencies. Moreover, system monitoring and control from a distance is challenging in the swiftly opportunistic IoT environment.



Edge computing introduces new computational layers atop the network infrastructure, between the cloud and the IoT devices, to where virtualized application resources can be moved. In close proximity to the devices, edge platforms have the capabilities to address the dynamics of the environment and serve local computation- and data-intensive applications with low latency and high bandwidth. Further, data processing at the edge layer increases data privacy and reduces load in the networks towards the cloud. 

Such edge platforms have already been presented \cite{Yousefpour2019, dolui2017, Bilal2018}. i) Cloudlets \cite{saty2009}, i.e. small-scale data centers, host mobile applications as virtual machines at the edge layer, ii) Multi-access Edge Computing (MEC) \cite{ETSI_SwForMEC_whitepaper}, previously known as Mobile Edge Computing, reuses network infrastructure components, such as cellular network base stations and Wi-Fi access points (AP), for data storage and hosting mobile applications, and iii) Fog computing \cite{bonomi2012} harnesses local networked devices, e.g. Wi-Fi APs and gateways but also laptops and set-top boxes, to form virtualized hierarchical computing platforms.


However, resource provisioning in the edge introduces pragmatic complexities. For example, network operators need to consider the placement of physical edge platform components such as servers. Often, constrained deployment budgets result in a limited number of servers with varying capacity and hardware. 
One option is to share server and network resources across the deployment, but a trade-off is that latencies are introduced and network load increased.
Furthermore, some locations, for example commercial centers or research facilities, may have deployment priorities over other locations. Finally, the deployment area may cover not only densely populated areas such as city centres, but also sparsely populated suburban areas. Without systematic approaches, the only way to solve these pragmatic issues is to rely on the domain expertise of the network operator.

Having decided on an initial server placement scheme, the next challenge is the online allocation and balancing of edge application workloads across the deployment. These two connected problems have been typically handled in steps, where workloads are allocated and balanced atop the initial server placement scheme, e.g. \cite{Kang19,xu2017}. The parameters for server placement are often derived from the observed online workload. Some works also focus solely on optimizing the online workload, e.g. \cite{yang2019,zhou2019,sharma2019,osanaiye2017,machen2017}. In such cases, the design of the underlying system is a significant factor, since the initially selected deployment scheme affects system performance in response to the online workload and latencies \cite{wang2020}.  The initial deployment decisions are thus pivotal in managing mobile applications across the multi-tenant edge platform to provide a high Quality of Service (QoS) for the edge platform and Quality of Experience (QoE) for mobile users. 

\begin{figure}[ht!]
\centering
 \includegraphics[width=0.4\linewidth]{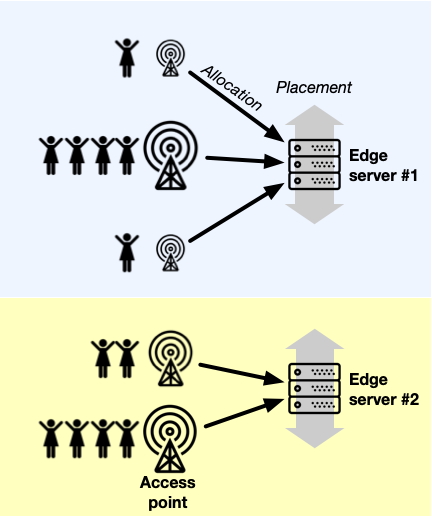}
 \caption{Edge server placement and workload allocation. Placement finds the optimal location for the edge servers, while user workloads, propagating through the APs, are allocated on the edge servers.}
 \label{fig:placement}
\end{figure}

An edge server placement scheme, which aims for maximal on-line QoS and QoE, involves two interrelated problems: first, the placement of edge servers, and second, the allocation of computing capacity to those servers to suit their expected workload (Figure \ref{fig:placement}). In this paper, we focus on solving these two problems optimally. First, we provide a survey of existing solutions for physical edge server placement and allocation, and identify gaps in those solutions. Second, based on the gaps identified, we develop a novel placement and allocation algorithm, called PACK, which takes into account an extensive set of parameters, including latencies, workload balancing and sharing, server co-location with APs, and prioritized locations. The resulting deployments are capable of handling the maximum workloads observed in the system while reaching both minimum latencies between the APs and the edge servers as well as a balanced load between the edge servers. The deployments thus ensure fair QoS across the deployment, and distribute application workloads such that the server capacities are respected, with minimal over- or under-provision. To increase the flexibility of the deployment, the workloads of APs can be shared between servers, addressing the  problem of overloaded low capacity servers. Finally, the algorithm allows preferring certain server locations over others, and providing reserve computational capacity as a reliability measure for edge application execution. 

These considerations lead to an NP-hard complex optimization problem which seeks for the best compromise between latency and workload balance. We consider this placement problem as an instance of the generic location-allocation framework \cite{farahani2009}. We adapt the framework for both MEC and Fog edge platforms, with high and low capacity servers respectively, and evaluate the resulting placement schemes with a set of performance statistics. The schemes are developed with a real-world public Wi-Fi network dataset \cite{kostakos2013} that consists of user equipment (UE) connections in both urban and suburban areas.

The contributions of this article are:
\begin{itemize}
    \item A detailed survey of existing physical edge server placement solutions. An extensive set of required parameters are collected and resulting gaps identified in the optimization of server deployments.
\item A location-allocation algorithm to provide optimal edge server placement with clustering of APs, based on the total system workload and number of servers, that minimizes the geospatial distances in the system, while considering server capacity constraints, workload balancing and sharing, reserve capacity and location preference. These additional constraints demonstrate real-world challenges and options faced in planning of edge deployments, which we consider as parts of critical edge infrastructure.
    \item A study of different edge deployment scenarios, i.e. MEC and Fog, in relation to the above set of parameters with a set of proximity statistics, to evaluate QoS between the presented algorithm and solutions in previous work.
    \item Development and evaluation of the placement algorithm, based on a massive-scale real-world data set ($\sim$257M data points) in a city-wide public Wi-Fi network, covering an area of a whole city with both dense and sparse AP deployments.     
    \item An open source software package\footnote{https://github.com/terolahderanta/rpack}, implemented in R, of the algorithm.
\end{itemize}

The rest of the paper is organized as follows. In Section \ref{background} we introduce and discuss the state-of-the-art in edge server placement, as well as the identified gaps in the research. In Section \ref{method} we present our algorithm, which we evaluate in Section \ref{evaluation}. We utilize a real-world data set in different scenarios and compare the results to the schemes based on previous work. Lastly, in Section \ref{discussion} we discuss the results and, in Section \ref{conclusion}, we present the conclusions. 


%% file: 2_related_work.tex
\section{Background}
\label{background}

Previous studies address edge server placement from four perspectives: (1) minimizing latencies, which are approximated using a number of different distance measures; (2) minimizing the deployment costs of the servers while limiting maximal latency; (3) optimizing the trade-off between latency and deployment costs; or (4) maximizing the user connections, i.e. coverage, within the clusters. The studies consider different sets of functional parameters for placement, such as the number of servers, individual server capacity, geo-locations of servers, and location priorities. Also, non-functional properties of QoS and QoE such as reliability have been used as parameters in the studied placement algorithms.

The studies applied varying approaches for placement: k-means \cite{liu2018}, k-means with mixed-integer quadratic programming \cite{guo2019}, density-based clustering \cite{Jia2017}, graph theory as in a minimum dominating set problem \cite{zeng2019}, hierarchical tree-like structures \cite{sinky2019,jiao2017}, multi-objective constraint optimization \cite{wang2019,bhatta2019, daSilva19}, mixed integer linear programming \cite{bouet2018, Fan17}, integer linear programming with heuristic solutions \cite{yao2016,Leyva19,Xu16}, non-linear mixed integer programming \cite{mondal2018}, clustering combined with optimization as a facility location problem \cite{mohan2018}, and with heuristic algorithms \cite{ma2017,yin2017}.

In comparison, PACK places a fixed number of servers while minimizing the latencies between users and edge servers, balancing the system workload and satisfying the lower and upper limits of server capacity. PACK can be seen as a variant of k-means type clustering problem with capacity constraints, solved here with a block coordinate descent algorithm with integer programming steps.

\subsection{Survey of edge server placement approaches}
\label{rel_work}

Next, we present a detailed survey of the existing solutions for physical edge server placement and analyze the set of parameters utilized in the algorithms. Further, we discuss the properties of our novel placement algorithm, PACK, presented in Section \ref{method}, addressing the identified gaps. The overall properties of the solutions are summarized in Table~\ref{rel_work_table}.

\subsubsection{Workload}\label{workload_sec}
Workload refers to the computing burden expected on the edge servers once deployed, resulting from online user requests. Most surveyed studies considered the total, average and/or worst-case historical workloads as proxies for the expected workload. The workloads were derived or estimated from user request data in varying granularity, taking into account, for example, the varying processing burdens of individual user requests, their total connection time, the duration of phone calls, the request rate of an AP, or the total number of requests to an AP. Worst-case expected workload was estimated by the maximum number of simultaneously connected users found in historical data \cite{yin2017, Guan19}. 


\hl{PACK considers workload by way of the setting a weight for each AP, reflecting the workload expected to accumulate on the edge servers through that AP. The higher the weight of an AP, the more influence it has in the placement. The selection of the weight value for each AP is not constrained; it is possible to use, for example, the historical peak loads or the average load values of the APs, or their request rates, perhaps scaled with a trend coefficient.

In our examples, we use the peak load, that is, the maximum number of connected users per AP in historical data, as the weights. With peak loads as weight values, we ensure sufficient server capacity in worst-case scenarios, e.g. sporadic peak hours.}
\subsubsection{Computing capacity} \label{sec:capacity}
\hl{Computing capacity refers to the computational and data resources of individual edge servers, available for the workload accumulating from the APs. Deciding upon capacities requires considering a trade-off between balanced workload and low latency. Indeed, minimizing the latency between APs and edge servers, the resulting deployment may have high workloads in busy areas and low workloads in quiet areas, leading to either unused capacity or edge servers with varying capacities. Conversely, balanced workloads, distributed evenly among edge servers, may lead to long distances between AP's and edge servers, manifesting as increased latencies.}


In related works, where a predefined number of servers were placed, typically either capacity limits, workload balance, or both were neglected. In some works, however, the servers had varying capacities. In more detail, most studies addressed computing capacity by prohibiting over-provisioning with a hard upper constraint on server workload. Alternatively, workload was balanced by distributing workload across the deployment \cite{guo2019, Jia2017, wang2019}, server capacity was scaled up on-demand regardless of the resulting cluster size \cite{ liu2018, zeng2019,sinky2019,yao2016,mohan2018}, or neighboring server capacity was utilized as a reserve or duplicated capacity \cite{Leyva19, Chen2018}. As a catch-all solution, if server capacity could not match its workload, the remaining load was discarded or offloaded to a cloud \cite{ Jia2017,daSilva19,bouet2018,   mondal2018,mohan2018, meng2017}, where resources were assumed to be unlimited.

A further alternative, cloudlet technology, allows scaling up server capacity as a backup measure to adapting to changing online workloads \cite{daSilva19,mondal2018,meng2017}. However, an initial cloudlet placement is still required, where additional capacity increases deployment and maintenance costs of the operator. A related solution is to support already deployed servers by adding new ones atop the existing deployment \cite{Loven2020a}. 

\hl{
Server placement schemes have mostly considered densely populated central areas (e.g.
} 
\cite{ Jia2017,wang2019,Fan17, mohan2018, Gedeon2018, Cui2020} 
\hl{
), but in some works also suburban areas were included (e.g. 
}
\cite{Jia2017,Fan17,Leyva19, mondal2018, mohan2018, yin2017}
\hl{). The trade-off between capacity and latency issue is exacerbated if the infrastructure has a high contrast between busy central areas and less busy suburban areas.
}


PACK supports both minimizing the latencies as well as balancing the workloads, ensuring sufficient server capacities, by applying both upper and lower capacity limits. Using both limits allows strong control over the latency–workload trade-off. For example, by prioritizing balance, one can set the capacity limits close to each other around the average workload. Alternatively, with sufficiently wide capacity limits, the lower limit can represent the capacity of one small server, and the upper limit the capacity of a large server. 

Further, a cloudlet can be built based on several low capacity servers. Thus, a wide interval allows flexibility with deployment areas with high variability in workloads, and more granularity in individual server capacity.

\subsubsection{Performance metrics} The performance of an edge server or the deployment is typically measured though a set of QoS and QoE parameters. ETSI standardization considers multiple metrics for demonstrating the performance deployments with both functional and non-functional sets of  Key Performance Indicators (KPIs) \cite{etsi006}, where the use of metrics depends on the use case. Functional metrics include latency, throughput, packet delivery and energy efficiency. Non-functional metrics include service availability, reliability and fault tolerance. 

Most of the studies measured performance by the latency between servers and users and/or APs, using different approaches. First, if the number of deployed servers was fixed, the average or total latency between APs and servers was used \cite{liu2018, guo2019, Jia2017, sinky2019,jiao2017, wang2019,yao2016,Xu16,ma2017, Chen2018, LiBo2018, Cui2020}. Additionally, the standard deviation of the server-wise workloads was used to evaluate load balancing \cite{guo2019, wang2019}. Second, if the aim was to minimize the number of servers or their energy consumption, the minimization was constrained by latency thresholds \cite{li2018,zeng2019, mohan2018, mondal2018}. 

Other performance measures in the studies considered both QoE and QoS aspects, including the percentage of satisfied user requests and the total utilization of the server capacity \cite{mohan2018}, the percentage of users within the expected distance \cite{yin2017}, the score and the cost of the deployment \cite{Leyva19}, intra-cluster traffic \cite{mohan2018}, energy consumption and average resource utilization \cite{li2018}, and the ratio of how much user demand can be allocated to cloudlets~\cite{Gedeon2018,daSilva19}.

Surveyed works evaluate the effect of various parameters to the QoS. First, the effect of the number of servers to latency \cite{liu2018, guo2019,Jia2017, sinky2019,jiao2017, wang2019,Xu16}, to workload balance \cite{guo2019, wang2019} and to hosted workload \cite{daSilva19} have been evaluated. Second, the effect of the capacity constraints to intra-cluster traffic, and of temporal changes to workload balance \cite{bouet2018}, as well as the effect of the number of APs to average access delay 
\cite{Xu16} and to the energy consumption and average resource utilization \cite{li2018} have been evaluated. Third, the link between other parameters and deployment budget has been evaluated, including the effect of distance thresholds \cite{zeng2019,Guan19, li2018} or capacity constraints \cite{zeng2019, bouet2018, Gedeon2018} to the number of servers. Further, Yin et al. \cite{yin2017} modelled the cost of the deployment as a function of the percentage of people within a given distance from the server, while Gedeon et al. \cite{Gedeon2018} considered the fixed costs of deploying an edge server, as well as the variable costs of using them based on the workload. Finally, Fan et al. evaluated deployment cost and latency as a function of the parameter that controls the trade-off between budget and proximity \cite{Fan17}.

\hl{
In summary, the surveyed articles measure deployment performance typically either (1) as the average QoS or (2) through setting a threshold for the worst tolerable QoS. However, our approach is to use multiple simultaneous performance measures, providing a clearer view of the resulting QoE and QoS. In more detail, we measure both the average and the worst-case latencies between the APs and servers, providing insights into the QoE across the whole coverage of the deployment. 
Further, we measure the resulting workloads on the edge servers, providing insights into the QoS of the resulting deployment.}

\subsubsection{Latency}


The communication latency between the APs and the edge servers is one of the main KPIs. Latency in an edge infrastructure, typically within a metropolitan area network, results from the architecture of the core network, i.e. its topology, and link capacities between the nodes. However, in all surveyed works the core network topology or the link capacities were not available. 
As a result, latency was approximated using computationally light proxies, such as topological or geophysical distances, depending on what kind of information was available.


With simulated topology \cite{guo2019, zeng2019,sinky2019,mohan2018,ma2017, jiao2017,Jia2017, Fan17, meng2017, yao2016, mondal2018, Chen2018, Xu16}, latency was either approximated by hop counts, or no specific latency measure was used at all. Alternatively, latency was appromated by the geospatial distances between the APs and the edge servers \cite{liu2018, wang2019, bouet2018, yin2017, li2018, Leyva19, LiBo2018, daSilva19, Guan19}. However, this requires that the coordinate space is consistent with respect to distance ranking \cite{yin2017}. Most studies used Euclidean distances, while some \cite{liu2018, LiBo2018} relied on squared Euclidean distances. Further, Wang et al. \cite{wang2019} scaled the Euclidean distances with respect to the density of the access points, while Leyva-Pupo \cite{Leyva19} et al. discretized the distances on multiple levels.


\hl{PACK is agnostic to the distance measure that represents latency, as discussed in detail in Section} \ref{sect_distance}\hl{. In more detail, PACK accepts any pairwise distance between the possible server locations and the APs.}

\hl{Like with all surveyed studies, the underlying physical network topology is not available in the data we use for evaluation }\cite{kostakos2013}\hl{. Hence, evaluating PACK, we approximate latency with the geospatial distances between APs and servers. To provide fair QoS across the deployment, we use squared Euclidean distances, as they have a tendency to produce spherical clusters with centralized heads} \cite{Negreiros06}. \hl{The use of squared distances results in a star-like topology with spatially centralized servers, contributing towards better worst-case proximity across the deployment, that is, lower latencies, regardless of the topological and physical positions of APs in the network.}
\subsubsection{Number of edge servers} 

The number of edge servers in the deployment is a trade-off between cost (e.g. operator budget and resulting available server capacity) and performance in relation to the user requirements. 

The surveyed studies relied on three main approaches to optimize the trade-off. First, based on a budget, a fixed number of servers were placed in a way that the latencies between the APs and the servers were minimized \cite{liu2018, guo2019,Jia2017, sinky2019,jiao2017,Xu16,wang2019, ma2017, Chen2018, meng2017, Gedeon2018, Cui2020, LiBo2018}. Second, the highest tolerated latency was defined and the number of servers was minimized with such latency constraint for each AP \cite{zeng2019,mohan2018,yao2016, mondal2018}. Third approach was to optimize the trade-off between the cost and the QoE, as number of satisfied users \cite{yin2017, daSilva19}, or the QoS of the resulting deployment \cite{Fan17, Leyva19, bhatta2019, Chen2018, Guan19}. Other approaches included  determining the number of servers as the result of minimizing energy consumption \cite{li2018} or maximizing connections within server regions \cite{bouet2018}.

In addition to the above trade-off, the deployment area needs to be considered. With city-wide real-world data sets, as in Section \ref{sec:dataset}, some deployments cover not only densely populated central areas, but also sparsely populated suburban areas, both with different requirements for server capacity. In such a deployment, minimizing the number of servers while satisfying latency constraints is no longer feasible, as in sparse areas the distances between servers and APs vary significantly, while remaining short in the denser areas. Therefore, such a placement scheme could lead to a large number of servers with a small number of connected APs.



\hl{Acknowledging that the network operator is often operating under a strict budget, PACK follows the majority of the surveyed studies and assumes a fixed number of servers. We evaluate PACK with the above-mentioned heterogeneous city-wide data set, covering both dense and sparse areas. As discussed by Wang et al.} \cite{wang2020}\hl{, the deployment scheme of the servers must be considered. Indeed, when communication latency can be relatively high, sparse servers in a flat architecture are preferred. However, if latency must be small and workloads are large, a hierarchical deployment architecture can increase edge system performance. Considering both viewpoints, we evaluate PACK in two different scenarios, named MEC and Fog, the former with a small number of high-capacity servers and the latter with a large number of low-capacity servers.}

\subsubsection{Co-location of servers and APs} \label{colocation_sec}

Deployment and maintenance costs of network operators are lower, and no new infrastructure is needed, when edge servers are co-located with APs. Most surveyed studies followed this constraint. As exceptions, co-location was not considered by some studies \cite{bouet2018,liu2018,yin2017}, while Bouet et al. considered spatial server areas, without exact server locations \cite{bouet2018}. Finally, Liu et al. \cite{liu2018} allowed the servers to be placed anywhere, and Yin et al. \cite{yin2017} considered new, possibly unforeseen, server locations. 

PACK follows the majority of the studies, co-locating edge servers with APs.



\subsubsection{Workload allocation} \label{sec:allocation}
The expected system workload, originating from user requests, propagates through the network from APs to edge servers. Most of the surveyed studies allocated the workload passing through individual APs to exactly one server, with some exceptions. Bouet at al. left some APs unallocated, without an edge server, as the overall workload of the servers exceeded their total capacity \cite{bouet2018}. Jiao et al. \cite{jiao2017} placed a fixed number of servers in a tree-structure, leaving some APs unassociated. With cloudlets, the workload of user requests was delivered to available cloudlets across the deployment \cite{Chen2018, Gedeon2018, LiBo2018, meng2017, ma2017, yao2016, Kang19, daSilva19,Xu16}. As a last resort, excessive workload was offloaded to the cloud. Da Silva et al. consider two types of workload, strict which can only be processed in a server and flexible that can be either hosted in the server or in the cloud \citep{daSilva19}. 

Fan et al. \cite{Fan17}, on the other hand, shared AP workload among several servers. In more detail, when the total workload on a server exceeds its capacity, workload must be shared between servers if offloading to cloud is not feasible. Particularly, this is beneficial with widely available low-capacity servers, e.g. laptops, as exemplified by Fog computing, where small-scale workload from local user requests can be handled in the premises. Shared workload can also relieve the trade-off between latency and workload balance, and improve the scalability of the placement algorithm. On a downside, whenever server capacity is available in premises, sharing complicates the management of the deployment and increases core network load. However, the studies did not consider the effects of both strict allocation and sharing of workload simultaneously. 


Adopting clustering terminology, we refer to workload sharing as \textit{fractional membership} and non-shared workload as \textit{hard membership}. PACK supports selecting the membership type depending on the scenario. Thus, both flat and hierarchical architectures \cite{wang2020}, e.g. MEC and Fog, hosting applications with different latency and workload requirements, are considered in placement.

\subsubsection{Location preference} \label{sec:locationpriority}


\hl{
Based on the business model or domain expertise of the network operator, city centres and other densely populated areas are often initially considered as edge server locations. Moreover, domain expertise and practical considerations, e.g., to guarantee a low latency to the edge server, may dictate a set of preferred locations, e.g. an airport, a commercial centre, or a research facility, even when the location could not otherwise be justified. Moreover, in a smaller scale, site-specific requirements, e.g. privacy concerns in a Fog deployment, may dictate placement of servers within a facility. Such fixed placement candidates can result in sub-optimal performance of the deployment, particularly with a limited budget. Further, as a practical handicap, network topology may constrain the locations in which the servers can be placed. In the surveyed studies, Leyva-Pupo et al. addressed location preference with site-specific latency requirements for the APs, and by reserving duplicate capacity for servers at prioritized APs }\cite{Leyva19}\hl{. Yao et al., on the other hand, placed a set of servers according to combined user mobility patterns }\cite{yao2016}\hl{.} 


PACK considers preferred edge server locations, based e.g. on domain expertise or practical issues, in both placement and allocation. K-means type clustering approaches are weight sensitive, that is, points with large weights attract a cluster center more strongly than light weighted points \cite{Ackerman12}. In our approach, the total weight of an AP is the sum of its workload and the location preference value for that AP, giving the preferred locations reduced latency in the placement. Thus, location preference is supported in the algorithmic level by adding a predefined constant to the workloads of the most critical APs. This approach further allows the use of several preference classes, or even a preference variable, where the size of the constant depends on the value of the variable.

\subsubsection{Reliability} \label{sec:reliability}

A system that is reliable and robust performs as designed and required for extended periods of time, and can resist and recover from failures. Shi et al. \cite{shi2016} discuss reliability challenges for edge infrastructure, where component failures can cause network and service interruptions. Further, user mobility and intermittent connectivity makes it challenging to provide a service and maintain QoE. Therefore, system-level management and recovery of failures is a crucial concern. Indeed, ETSI considers reliability as one of the non-functional KPIs for edge services \cite{etsi006}. 

In the surveyed studies, Leyva et al. \cite{Leyva19} address reliability in workload allocation by providing backup capacity in servers for every AP. This way, if a server fails, a replacement is readily available. A drawback is that the deployment costs increase and possibly unused capacity will be reserved.

Further, Cui et al. \citep{Cui2020} address the concern of runtime failures in edge server networks by optimizing the placement of edge servers based on a trade-off between user coverage and network robustness. They form a placement strategy with weight parameters for both user coverage and robustness that represent the preference between those two parameters.

\hl{
To address reliability, PACK allows the workload of APs to be replicated and assigned to any number (typically two) of servers instead of just one. Minding budget constraints, the decision on workload replication can be made for each AP individually, selecting only those APs that are deemed as critical network infrastructure by a domain expert. 
}


\subsubsection{Data sets} 

Either simulated or real-world data sets were used for the development and evaluation of the placement algorithms. All the presented data sets were ca. five years old, even those used in the more recent studies. To the best of our knowledge, data sets of large-scale edge computing infrastructures do not exist yet. 

Multiple studies used simulated data sets \cite{zeng2019, jiao2017, Leyva19,Fan17,ma2017,yao2016, Chen2018, LiBo2018, mondal2018, Guan19}, with the number of APs varying from tens to hundreds. The simulation was sometimes based on additional data such as city population data \cite{Jia2017,Xu16, sinky2019, bhatta2019} or the locations of cell towers, public Wi-Fi APs and street lamps, as well as user mobility traces \cite{Gedeon2018}.

A number of studies based their evaluations on real-world Wi-Fi or mobile network data sets. The geo-referenced phone call records in Milan, Italy, from November 2013 to January 2014 \cite{barlacchi2015} was the most popular \cite{liu2018,bouet2018, mohan2018,daSilva19}, while a Shanghai Telecom data set with 4.6 million call records and 7.5 million movement traces of 10 thousand mobile users in 3000 base stations, over a period of six months in 2014 \cite{wang2017qos}, came a close second \cite{guo2019,wang2019,li2018}. Yin et al. obtained their data from the globally-distributed PlanetLab nodes as well as measurement nodes deployed in China \cite{yin2017}. Guan et al. utilized the traces of 320 taxi cabs in Rome, Italy, during one month in 2014 \cite{Guan19}. 

We evaluate PACK with a real-world data set collected across a city-wide open Wi-Fi network deployment in Oulu, Finland. The data set consists of 450 Wi-Fi APs, deployed into both central and suburban areas, and all of their connected users during the year 2014 ($\sim$257M data points). The data set is described in detail in Section \ref{sec:dataset}. 
\subsubsection{Algorithmic scalability} 

Algorithmic scalability refers to how the running time of the algorithm depends on the data. In general, the better results the algorithm provides, the worse it scales. Some studies presented heuristic algorithms \cite{yin2017,Xu16,Chen2018,ma2017}, that have good scalability but with no guarantees on result optimality. Most of the studies relied on formal optimization methods which, while more complex, guarantee that the result is at least locally optimal. An exception, Ma et al. propose particle swarm optimization to address the scalability of an initially heuristic algorithm \cite{ma2017}.  

For formal optimization methods, the fewer constraints in the optimization, the less control the user has but the better the algorithm scales. The simplest optimization methods had no capacity constraints \cite{liu2018,sinky2019, Kang19}, being scalable but with the least control. The studies that included optimization constraints improved scalability with an iterative approach where location and allocation steps alternated in an iterative optimization algorithm. Leyva-Pupo et al. \cite{Leyva19} used for large-scale data sets variants of hybrid simulated annealing and evolutionary algorithms to alternatively optimize placement and allocation with capacity constraints.

Some studies first partitioned the workload of APs into clusters, without applying capacity constraints, after which the servers were placed at each cluster separately \cite{mohan2018,yin2017}. Alternatively, servers were first placed, without workload balancing, and then APs allocated to servers with the aim to balance the workload \cite{guo2019, Jia2017}. Further, Guo et al. \cite{guo2019} used quadratic optimization for the balanced allocation step, thus reducing computation time but consuming memory as the whole data is allocated at once.  Jia et al. relied on a hybrid model, using a fast but heuristic approach for the balanced allocation step \cite{Jia2017}. Also, Guan et al. utilize a two-step method where the area was first partitioned based on minimizing the service migration costs, and then the location of servers and the allocation of APs were optimized separately \cite{Guan19}.

Bouet et al. \cite{bouet2018} set a dense grid in the designated area, and then merged the grid cells by their workloads to obtain the spatial extents of the servers. The computational time thus depended on the number of cells and not on the number of APs. Thus, the method scales well in relation to the number of APs, but not with respect to the spatial extent.




\hl{PACK relies on formal optimization, and we improve the scalability of our approach by a block-coordinate descent algorithm with integer programming steps. The chance of local optima are reduced by repeating the optimization with a large number of initial values. We also allow fractional membership, improving scalability. The scalability of our approach depends on the number of APs but not on the spatial extent of the deployment coverage. This is important as we focus on large-scale infrastructure with both busy city centres and quiet suburban areas.} 

\subsubsection{Source code} 
\hl{None of the surveyed studies has published the source code for their methods.}
PACK source code is published as an open source project, called \textit{rpack}, in GitHub \cite{rpack}. The algorithm is implemented with R and the package contains all the described functionality with an easy-to-use API.


\begin{landscape}
\begin{table*}[t]

\footnotesize

\centering
{\renewcommand{\arraystretch}{1}
 \begin{tabular}{l l l l l l l  } 
 \toprule
 \textbf{Article} 
 & \textbf{Workload}   
 & \begin{tabular}{@{}l@{}}\textbf{Capacity} \\ \textbf{constraints}\end{tabular} 
 & \begin{tabular}{@{}l@{}}\textbf{Distance} \\ \textbf{Measure}\end{tabular} 
 & \begin{tabular}{@{}l@{}}\textbf{Role of} \\  \textbf{distance}\end{tabular} 
 & \begin{tabular}{@{}l@{}}\textbf{Number of} \\ \textbf{servers}\end{tabular} 
 &  \begin{tabular}{@{}l@{}}\textbf{Server} \\ \textbf{association}\end{tabular} \\
 \midrule 

Bhatta et al.\cite{bhatta2019} & workload & upper & netw. topology & thres. \& min. & minimized & hard 
\\
Bouet at al.\cite{bouet2018} & phone calls & upper & geospatial & not used & on-demand & hard 
\\	
Chen et al. \cite{Chen2018} & simulated & upper & netw. topology & thresh. \& min. & fixed & request 
\\
Cui et al. \cite{Cui2020} & total no. of requests & not used & geospatial & thresholded & fixed & hard 
\\
da Silva et al. \cite{daSilva19} & time series & threshold & geospatial & not directly used & on-demand & request 
\\
Fan et al.\cite{Fan17} & requests & upper & netw. topology & minimized & on-demand & fractional 
\\
Gedeon et al. \cite{Gedeon2018} & simulated & upper & netw. topology & thresholded & fixed & request 
\\
Guan et al. \cite{Guan19} & peak & threshold & geospatial & minimized & on-demand & hard  
\\ 
Guo et al.\cite{guo2019} & total requests & balanced & netw. topology & minimized & fixed & hard	
\\	
Jia et al.\cite{Jia2017} & request rate & balanced & netw. topology & thresholded & fixed & hard   
\\
Jiao et al.\cite{jiao2017} & workload & not used & netw. topology	& minimized	& fixed	& hard	
\\	
Kang et al.\cite{Kang19} & total requests &	not used  &	geospatial & minimized & fixed & request 
\\
Leyva-Pupo et al.\cite{Leyva19} & simulated & upper & geospatial & thres. and min. & on-demand & duplicate
\\ 
Li et al. \cite{LiBo2018} & simulated & upper & netw. topology & minimized & fixed & request 
\\
Li et al.\cite{li2018} & total conn. time & upper & geospatial & thresholded & on-demand & hard 
\\	
Liu et al.\cite{liu2018} & not used & not used & geospatial & minimized	& fixed	& hard		
\\	
Ma et al.\cite{ma2017} & request rate &	upper \& balanced &	netw.topology & minimized & fixed & request 
\\
Meng et al.\cite{meng2017} & total requests & upper & netw. topology & minimized & fixed & request 
\\	
Mohan et al.\cite{mohan2018} & average requests & not used & netw. topology &	thresholded	& minimized	& hard
\\
Mondal et al.\cite{mondal2018} & request rate & upper & netw. topology & thresholded & minimized & hard 
\\
Sinky et al.\cite{sinky2019} & total requests &	not used & netw. topology	& minimized	& fixed	& hard
\\
Wang et al.\cite{wang2019} & total requests & balanced & geospatial & minimized	& fixed	& hard 	
\\
Xu et al. \cite{Xu16} & total no. of requests & upper & netw. topology & minimized & fixed & request  
\\
Yao et al.\cite{yao2016} & total requests & upper & netw.  topology & minimized & on-demand & request 
\\
Yin et al.\cite{yin2017} & peak & upper & geospatial & thresholded & on-demand & hard 	
\\
Zeng et al.\cite{zeng2019} & workload & upper & netw. topology & thresholded & minimized & hard 
\\	
\textbf{This article} &	peak & upper \& lower & geospatial & minimized & fixed & hard or frac. 
\\

 \bottomrule
 \end{tabular}}
 \caption{Comparison of edge server placement algorithms with respect to their overall properties.
 }
 \label{rel_work_table}
\end{table*}
\end{landscape}

\subsection{Summary}

In summary, the PACK algorithm advances the previous studies as follows:
\begin{itemize}
    \item Placing a fixed number of edge servers, PACK does two things simultaneously. Latencies between APs and servers are minimized, and the workload of the edge servers is balanced, satisfying both lower and upper capacity constraints of the servers.
    \item PACK allows both wide and tight capacity ranges and sharing the workload of APs, leading to flexible and scalable deployment in different scenarios, depending on the requirements and server capacity. This is particularly beneficial in Fog computing with low-capacity servers.
    
    \item \hl{PACK allows designating APs as preferred, i.e., to be located close to an edge server. The preference of APs was largely omitted in the related work, except with site-specific latency requirements by Leyva et al. }\cite{Leyva19}\hl{.}
    
    \item PACK addresses reliability concerns in the deployment by providing reserve capacity for APs that are considered a part of critical network infrastructure. In the related work, double capacity for the whole deployment was considered by Leyva et al. \cite{Leyva19}, but otherwise omitted. 
    
    \item PACK is evaluated with latency statistics on both QoS and QoE, using sample quantiles, and including both average and worst case latencies.        
    
    \item \hl{The development and evaluation of PACK relies on a large-scale real-world data set, which includes both dense central and sparse suburban areas with wide range of workloads.} 
    
    \item \hl{PACK is evaluated on scenarios with both a low number of high-capacity edge servers ("MEC") and a high number of low-capacity servers ("Fog").}
    
       
\end{itemize}
In Table~\ref{reunaehto_table}  we compare how our approach compares to others related to the gaps identified.

\begin{table}[ht!]
    \centering
    \footnotesize
    \caption{Identified gaps in the related work.}
    \begin{tabular}{l l l l l l l l l l}
        Article     
        & \rot[60]{No latency constraints} 
        & \rot[60]{Upper capacity constraints}  
        & \rot[60]{Balanced workload} 
          & \rot[60]{Formal optimization}  
        & \rot[60]{Both memberships}  
        & \rot[60]{Real data} 
        & \rot[60]{Location preference} 
        & \rot[60]{Reliability}
        & \rot[60]{Source code} 
        \\
         \midrule
        Bhatta et al.\cite{bhatta2019}
        & -
        & \checkmark
        & -
        & -
        & -
        & \checkmark
        & -
        & -
        & -
        \\
        Bouet at al.\cite{bouet2018}
        & \checkmark
        & \checkmark
        & - 
        & - 
        & -
        & \checkmark
        & -
        & -
        & -
        \\
        Chen et al. \cite{Chen2018}
        & \checkmark
        & \checkmark
        & -
        & -
        & -
        & -
        & -
        & -
        & -
        \\
        Cui et al. \cite{Cui2020}
        & -
        & \checkmark
        & -
        & \checkmark
        & -
        & \checkmark
        & -
        & \checkmark
        & -
        \\
        da Silva et al. \cite{daSilva19}
        & -
        & \checkmark
        & -
        & -
        & -
        & \checkmark
        & -
        & -
        & -
        \\
        Fan et al.\cite{Fan17} 
        & \checkmark
        & \checkmark
        & -
        & \checkmark
        & -
        & -
        & -
        & -
        & -
        \\
        Gedeon et al. \cite{Gedeon2018}
        & \checkmark
        & \checkmark
        & -
        & -
        & -
        & \checkmark
        & -
        & -
        & -
        \\
        Guan et al. \cite{Guan19}
        &-
        & \checkmark
        & -
        & -
        & -
        &\checkmark
        & -
        & -
        & -
        \\
        Guo et al.\cite{guo2019}
        & \checkmark
        & -
        & \checkmark
        & -
        & -
        & \checkmark
        & -
        & -
        & - 
        \\
        Jia et al.\cite{Jia2017}
        & -
        & -
        & \checkmark
        & -
        & -
        & \checkmark
        & -
        & -
        & -
        \\
        Jiao et al.\cite{jiao2017}
        & \checkmark
        & -
        & - 
        & -  
        & - 
        & -
        & - 
        & -
        & - 
        \\
        Kang et al.\cite{Kang19}
        & \checkmark
        & -
        & \checkmark
        & -  
        & -
        & -
        & -
        & -
        & -
        \\
        Leyva-Pupo et al.\cite{Leyva19}
        & \checkmark
        & \checkmark
        & -
        & \checkmark
        & -
        & -
        & \checkmark
        & \checkmark
        & -
        \\
        Li et al. \cite{LiBo2018}
        & -
        & \checkmark
        & -
        & \checkmark
        & -
        & -
        & -
        & -
        & -
        \\
        Li et al.\cite{li2018}
        & -
        & \checkmark
        & -
        & -  
        & -
        & \checkmark
        & -
        & -
        & -
        \\
        Liu et al.\cite{liu2018}        
        &  \checkmark 
        &  -     
        &  -
        &  \checkmark 
        &  -
        &  \checkmark
        & - 
        & -
        & -
        \\
        Ma et al.\cite{ma2017}
        & \checkmark
        & \checkmark
        & \checkmark
        & -  
        & -
        & -
        & -
        & -
        & -
        \\
        Meng et al.\cite{meng2017}
        & \checkmark
        & \checkmark
        & -
        & -
        & -
        & -
        & -
        & -
        & -
        \\
        Mohan et al.\cite{mohan2018}
        & -
        & -
        & -
        & -
        & - 
        & \checkmark
        & -
        & -
        & -
        \\
        Mondal et al.\cite{mondal2018}
        & -
        & \checkmark
        & -
        & \checkmark
        & -
        & -
        & -
        & -
        & -
        \\
        Sinky et al.\cite{sinky2019}
        & \checkmark
        & -
        & -
        & -
        & -
        & \checkmark
        & -
        & - 
        & -
        \\
        Wang et al.\cite{wang2019}
        & \checkmark
        & -
        & \checkmark
        & \checkmark 
        & -
        & \checkmark
        & -
        & -
        & -
        \\
        Xu et al. \cite{Xu16}
        & \checkmark
        & \checkmark
        & -
        & -
        & -
        & -
        & -
        & -
        & -
        \\
        Yao et al.\cite{yao2016}
        & -
        & \checkmark
        & \checkmark
        & -  
        & -
        & -
        & -
        & -
        & -
        \\
        Yin et al.\cite{yin2017}
        & -
        & \checkmark
        & -
        & -
        & -
        & \checkmark
        & -
        & -
        & -
        \\
        Zeng et al.\cite{zeng2019}
        & -
        & \checkmark
        & - 
        & - 
        & -
        & -
        & -
        & -
        & - 
        \\
        \midrule
        \textbf{This article}
        & \checkmark
        & \checkmark
        & \checkmark 
        & \checkmark 
        & \checkmark 
        & \checkmark 
        & \checkmark 
        & \checkmark
        & \checkmark
        \\
    \bottomrule
    \end{tabular}
    \label{reunaehto_table}
\end{table}

%% file: 3_method.tex

\section{PACK algorithm for edge server placement}
\label{method}

In this section, we present in detail the PACK algorithm for physical edge server placement. The PACK algorithm optimizes the placement of a fixed number of edge servers, minimizing the distances between servers and APs while balancing system workload, and satisfying server capacity constraints with both lower and upper limits. The algorithm maintains fair QoS, regardless of the AP deployment density, in a dispersed city-wide data set. PACK considers QoS with a pairwise distance measure between APs and servers, such as latency or hop distance, if available, or geospatial distance. Further, PACK provides a number of parameters which model workload sharing, workload replication, and allow preferring certain locations over others. Finally, we consider the evaluation of PACK results by way of QoE and QoS metrics.

\subsection{The PACK-algorithm}
\label{pack}
 

\hl{
Given a data set of $n$ APs, let $x_i$ be the coordinates for AP $i$. Further, AP $i$ has workload $w_i$, which can be determined by, for example, the average or total amount of connected users per AP 
}
\hl{(see section} \ref{workload_sec}).


\hl{Furthermore, the total weight value $a_i$ for AP $i$ is defined as $a_i=w_i+\gamma_i$, where $\gamma_i$ is the location preference of the AP $i$, set for example by a domain expert (see section } \ref{sec:locationpriority} \hl{). The default value for $\gamma_i$ is 0, that is, no preference is set. For the preferred APs, we set the value of $\gamma_i > 0$. The value of $\gamma_i$ can be considered as artificial workload added to AP $i$. The higher the value of $\gamma_i$, the bigger the total weight of AP $i$ is in the optimization}


\hl{We place a fixed number $k$ of edge servers, aiming to minimize weighted distances to their associated APs and satisfying the capacity constraints. Such a placement scheme can be considered as a capacitated location allocation problem} \cite{farahani2009, Brimberg08, Cooper64}\hl{. Let us denote by $c_j$, $j = 1,2,...,k$, the coordinates for the  $k$  edge servers, each of which is assumed to have similar capacity for computational load.}

\hl{Denote by $y_{i j}$ the membership of the AP $i$ to the edge server $j$. In other words, $y_{i j}$ is an indicator variable of the allocation of the workload, where $y_{i j} = 1$ if the workload in AP $i$ is allocated to the server $j$ and $y_{i j} = 0$ otherwise. We address the reliability of the placement by assigning selected APs to more than one (say, two) servers (see section }\ref{sec:reliability}\hl{; cf. the Q-coverage problem} \cite{Karatas16}\hl{). Lastly, $d(i,j)$ denotes the distance between AP $i$ and server $j$. The algorithm allows the use of any pair-wise distance metric $d$, as discussed in section~}\ref{sect_distance}.



Our goal is to minimize the sum of weighted distances between the edge servers and the APs they cater for, while taking into consideration the workload of each AP and the capacity constraints of each server. Hence, we minimize the objective function

\begin{equation}
\label{eq:loss}
\argmin_{c_j, y_{i j}} \sum_{i = 1}^{n} \sum_{j = 1}^{k} a_i d(i,j) y_{i j}
\end{equation}
with the following four constraints: 
\begin{enumerate}
    
    \item $c_j \in \{x_1, x_2,... ,x_n\} \quad \forall j$. 
    \label{const1}
    
    Each edge server location $c_j$ is assumed to be the same as one of the AP locations $x_i$ \hl{(see section }\ref{colocation_sec}). 
    
    
    \item a) $y_{i j} \in \{0, 1\}$ or b) $y_{i j} \in [0,1] \quad \forall i,j$. \label{const2}
    
    \hl{Either hard (a) or fractional (b) membership constraints are used (see section }\ref{sec:allocation}\hl{). In the case of hard membership, the workload of an AP $i$ is either completely allocated to edge server $j$ ($y_{ij} = 1$), or not at all ($y_{ij} = 0$). In the case of fractional membership,  $y_{i j}$ represents the fraction of workload of AP $i$ assigned to 
    server $j$ }\cite{Borgwardt17}. 
    
    \item $\sum_{j = 1}^{k} y_{i j} = q_i \quad \forall i$. \label{const3}
    
    \hl{
    All the workload of an AP is assigned to $q_i$ servers, where $q_i>1$ for APs deemed critical by a domain expert, and  $q_i=1$ for others (see section }\ref{sec:reliability}\hl{). In case of hard membership, $q_i$ must be an integer. 
    }
    

    \item $L \leq \sum_{i = 1}^{n} w_i y_{i j} \leq U \quad \forall j$. \label{const4}
    
    \hl{Capacity for each edge server is constrained between $L$ and $U$ (see section }\ref{sec:capacity} \hl{).}
    
\end{enumerate}



\hl{
For solving the optimization problem, we use a block-coordinate descent algorithm }\cite{Tseng01}\hl{. In Algorithm }\ref{alt_alg}\hl{, we iterate two main steps: (i) the allocation of the APs to the given locations of servers (line }\ref{allocation_step}\hl{) and (ii) re-locating each server given the APs allocated (line }\ref{location_step}\hl{). The iteration is continued until the locations of the edge servers do not change.
}

As the result of such an iteration is not guaranteed to be the global minimum, $N$ initial values for the edge server locations are used in Algorithm~\ref{alt_alg}. A set of reasonable initial values is obtained using the k-means++ algorithm \cite{Arthur07}. The k-means++ algorithm spreads the initial server locations, which improves both the speed and the accuracy of the k-means method. 

\textbf{Allocation-step.}
In the allocation-step, we minimize the objective function (\ref{eq:loss}) with respect to $y_{i j}$. Only constraints \ref{const2}, \ref{const3} and \ref{const4} are used, since edge server locations $c_j$ are assumed to be fixed. For the hard membership constraint, this step is an NP-hard integer programming task \cite{papadimitriou1981}. However, in the case of the fractional membership, this is a linear programming task that can be solved in polynomial time \cite{khachiyan1980}.

\textbf{Location-step.}
Location-step relocates the edge servers based on the allocation of the APs. In other words, the objective function (\ref{eq:loss}) is minimized with respect to $c_j$  with only constraint \ref{const1}, while keeping the allocations $y_{i j}$ fixed. This step is identical to calculating for each server the medoid of the APs assigned to it, where medoid is defined as the point from which the sum  distances to all the points in the cluster is minimized \cite{Kaufmann87}.

\begin{algorithm}[t]
\caption{PACK-algorithm}
\textbf{Input:} $x_i, w_i, k, N$ \\
\textbf{Output:} Edge server locations $c_j^*$ and allocations $y_{i j}^*$\\
\label{alt_alg}
\begin{algorithmic}[1]
\FOR{$i = 1$ to $N$}
\STATE Initialize $c_j$, $j = 1,2,...,k$ using k-means++
\WHILE{$c_j$ changes}
\STATE Allocation-step: minimize (\ref{eq:loss}) with respect to $y_{i j}$ \label{allocation_step}
\STATE Location-step: minimize (\ref{eq:loss}) with respect to $c_j$ \label{location_step}
\STATE $S \leftarrow $ the value of the objective function
\ENDWHILE
\IF{$S < S_{min}$ or $i = 1$}
\STATE $S_{min} \leftarrow S$
\STATE $c_j^* \leftarrow c_j$
\STATE $y_{i j}^* \leftarrow y_{i j}$
\ENDIF
\ENDFOR
\RETURN $c_j^*, y_{i j}^*$
\end{algorithmic}
\end{algorithm}


We implemented the block-coordinate descent algorithm as an open source R package called \textit{rpack} \cite{rpack}. The allocation step is run on Gurobi \cite{gurobi}, a fast optimizer package with R bindings. If Gurobi is not available, the allocation-step can be run on the lpSolve package for R \cite{lpsolve}.


\subsection{Distance}
\label{sect_distance}

In the related work, typical choices for the distance measure include the Euclidean and squared Euclidean distances, 
\[
\begin{split}
d_1(i,j)&=|| x_i - c_j ||, \\
d_2(i,j)&=|| x_i - c_j ||^2,
\end{split}
\]
respectively.

In the case of Euclidean distances the problem can be considered as the capacitated p-median problem \cite{farahani2009} and the squared Euclidean distances as a capacitated k-means clustering, where the cluster centers are constrained to the data points. Such a discrete variant of k-means method is generally referred to as a k-medoid method \cite{Kaufmann87}. 

By applying the Euclidean distances in (\ref{eq:loss}), the average distances form the servers are minimized. However, the  minimization of the  squared Euclidean distances imposes a stronger penalty from large deviations, thus there is a tendency for more spherical-like clusters with centralized cluster heads \cite{Negreiros06}. Such a scheme results in a star-like deployment topology with spatially centralized edge servers. This contributes towards better latency, i.e. fair QoS for both dense and sparse AP deployments, being particularly beneficial for otherwise remote APs. Thus, we prefer the squared Euclidean distances and utilize them in our evaluations.  


\hl{
The PACK algorithm is agnostic towards the choice of the distance metric. By default, our open source \textit{rpack} R package allows the use of both Euclidean and squared Euclidean distances. However, providing \textit{rpack} with an $n \times n$ matrix of all pair-wise distances between APs, any distance metric can be applied. This makes it possible to use metrics derived from the network topology, such as hop counts or latency metrics, if available. Even more generally, one can optimize any quantitative KPI that can be defined as a sum of pairwise functions of the APs and the servers they are assigned to. An example are objective and subjective edge service-dependent metrics as presented in} \cite{etsi006}\hl{, including the type of service, either real-time and non-real-time, or responsiveness as an QoE metric, which set their requirements for the placement candidates.}

\subsection{Evaluation metrics}\label{eval_metrics}


\hl{
We evaluate PACK with a set of both functional and non-functional KPIs, related to the QoS and the workload accumulating on the servers. Specifically, we focus on the distribution of latency between APs and edge servers, and the distribution of workload on the edge servers.
}



We measure latency with the selected distance metric, i.e., the distances between APs and servers. However, the same QoS metric can be interpreted as a QoE metric from a user point-of-view, i.e., as latency measured by how close the users are to the servers, by way of the APs they are connected to. For the sake of interpretability, we apply the Euclidean distance as QoS metric as opposed to the squared Euclidean distance used in the placement.

The average AP distance weighted by the workload is obtained as
\[
\text{Mean}= \frac{1}{W}\sum_{i = 1}^{n} \sum_{j = 1}^{k} w_i || x_i - c^*_j || y^*_{i j},
\]
\hl{
where $c^*_j$ are the optimized server locations, $y^*_{i j}$ are the optimized memberships and $W=\sum_{i = 1}^{n}  w_i$ is the total workload of the area.
}


\hl{
Further, we explore the distance distribution thoroughly with the sample quantiles $q_\alpha$, which measure the maximal distance between AP and its edge server for $\alpha$ proportion of the workload. Hence, by selecting a  proportion $\alpha$ close to 1, the worst case QoS can be evaluated. For example, $q_{0.95}$ is the distance from 95\% of the workload to the server. Further, $\alpha=0.5$, $q_{0.5}$ corresponds to the median QoS.  
}


\hl{
Finally, we measure the workload balance of the placement candidates with the standard deviation of the edge server workloads, as well as the minimum and maximum workload values. 
}

 

%% file: 4_evaluation.tex

\section{Evaluation}
\label{evaluation}


\hl{To evaluate the PACK algorithm, we simulate edge server placement with a real-world Wi-Fi network data set in two edge computing scenarios: one with a small number of high-capacity edge servers (the MEC scenario), and the other with a large number of low-capacity edge servers (the Fog scenario). For the analysis, these two scenarios provide different real-world based viewpoints and validation criteria to evaluate both the flexibility of the PACK algorithm and the effects of the extensive set of parameters available in PACK. The aim of the evaluation is to provide insights into the resulting edge computing architecture, in terms of number of servers and their capacity ranges,} and system design factors such as scalability and costs in deployment and management, as discussed for example in \cite{wang2020}\hl{. Further, the evaluation considers a set of additional real-world based} constraints on server capacity and workload sharing, location priorities, and reliability by way of workload replication.

Both the MEC and Fog scenarios present a number of setups which highlight different aspects of placement. In all setups, PACK uses 100 initial placement candidates (random sets of server locations), selected with the k-means++ method. Final server placement and the allocation of APs to edge servers are based on clustering with the smallest value of the objective function (Equation \ref{eq:loss}), as described in the PACK algorithm (Algorithm \ref{alt_alg}). 

\subsection{Data}

\label{sec:dataset}

\begin{figure*}[ht!]
  \centering
\begin{subfigure}[t]{.47\textwidth}
  \centering
  \includegraphics[width=1\textwidth]{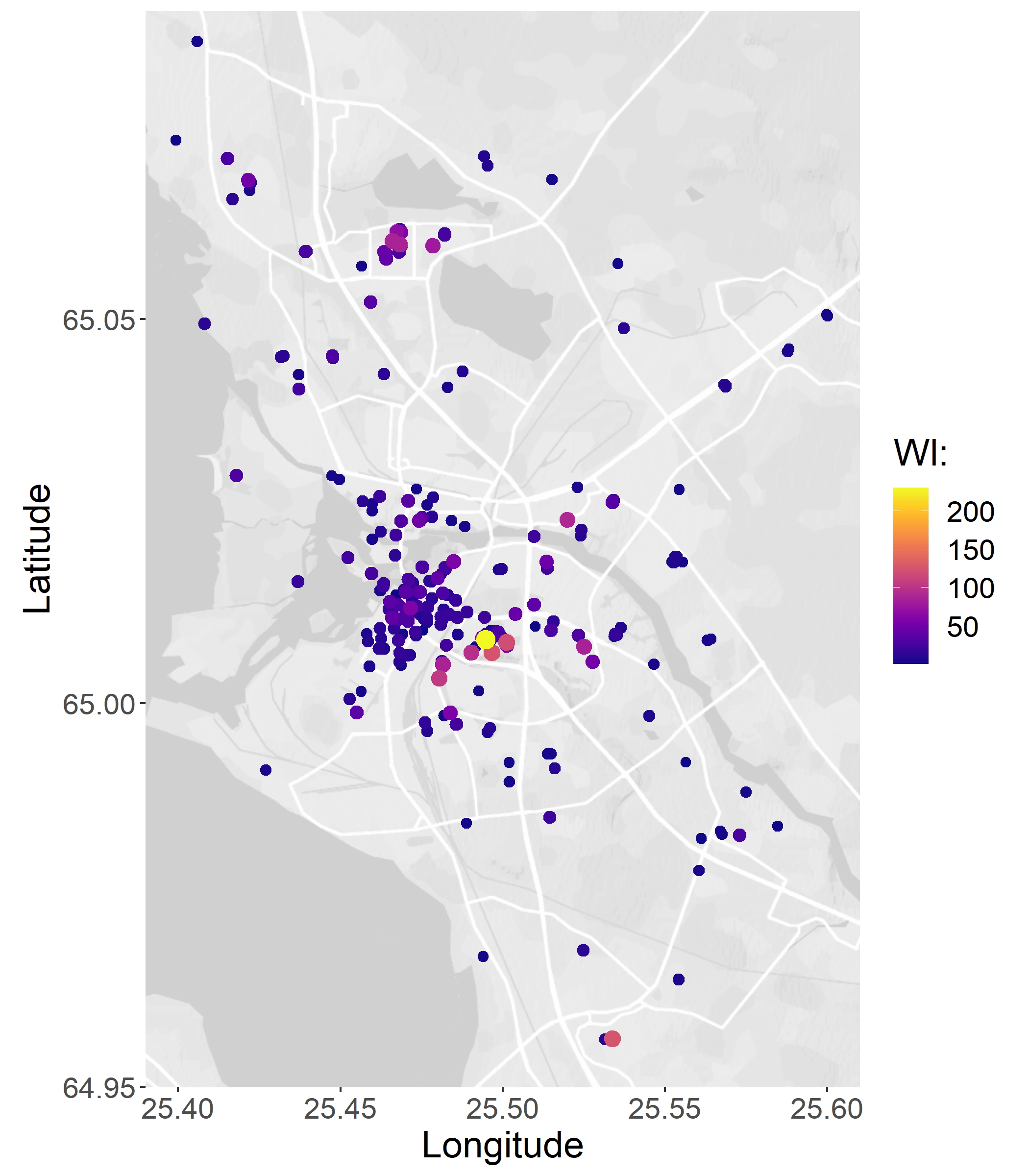} 
  \caption{PanOULU AP deployment. \\ Wl: workloads of the APs.}
  \label{bs_workloads}
\end{subfigure}
\begin{subfigure}[t]{.47\textwidth}
  \centering
  \includegraphics[width=1\linewidth]{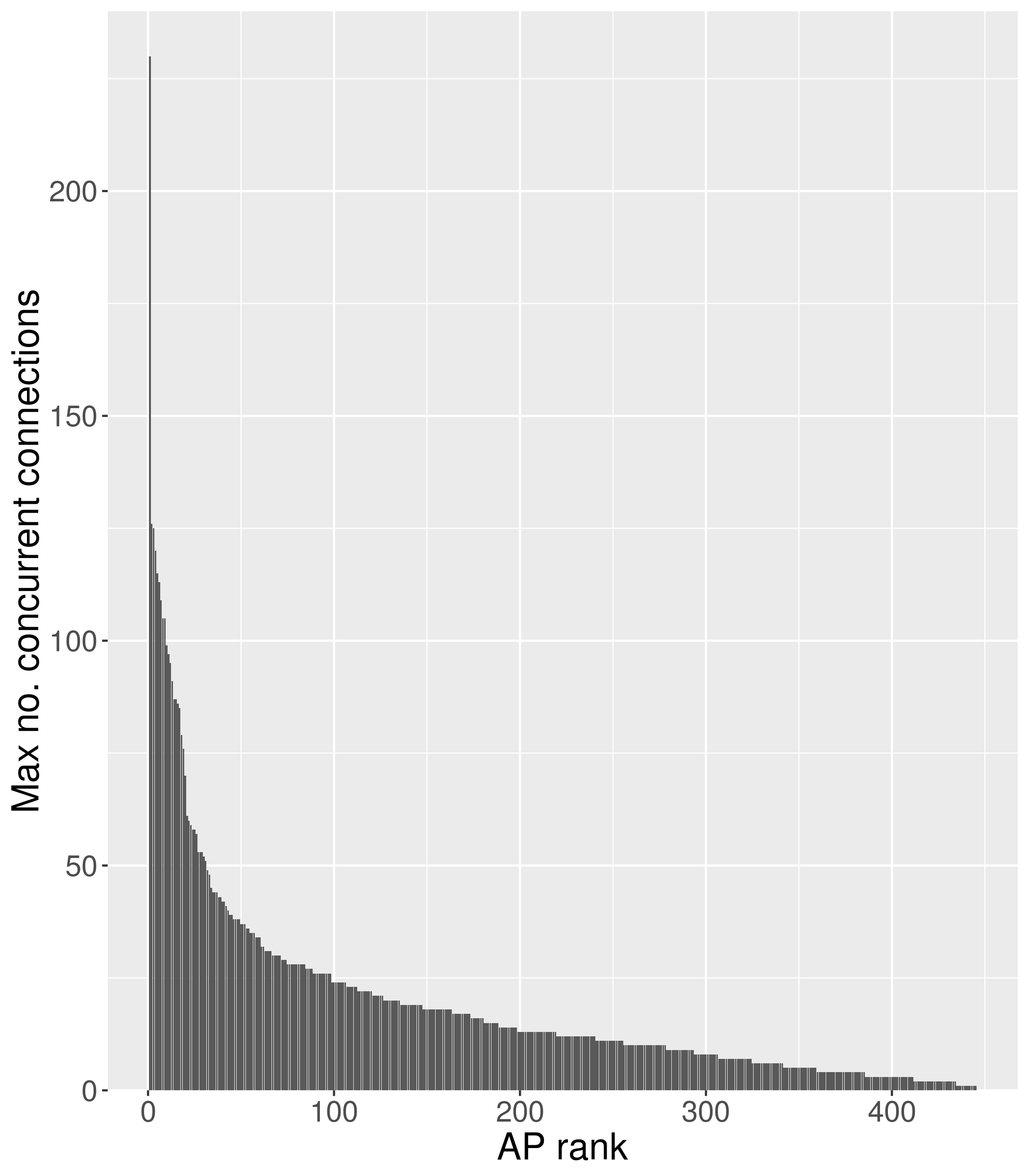}
  \caption{Maximum number of connected users on each AP. }
  \label{ap_map_hist}
\end{subfigure}
\begin{subfigure}[t]{0.6\textwidth}
  \centering
  \includegraphics[width=1\linewidth]{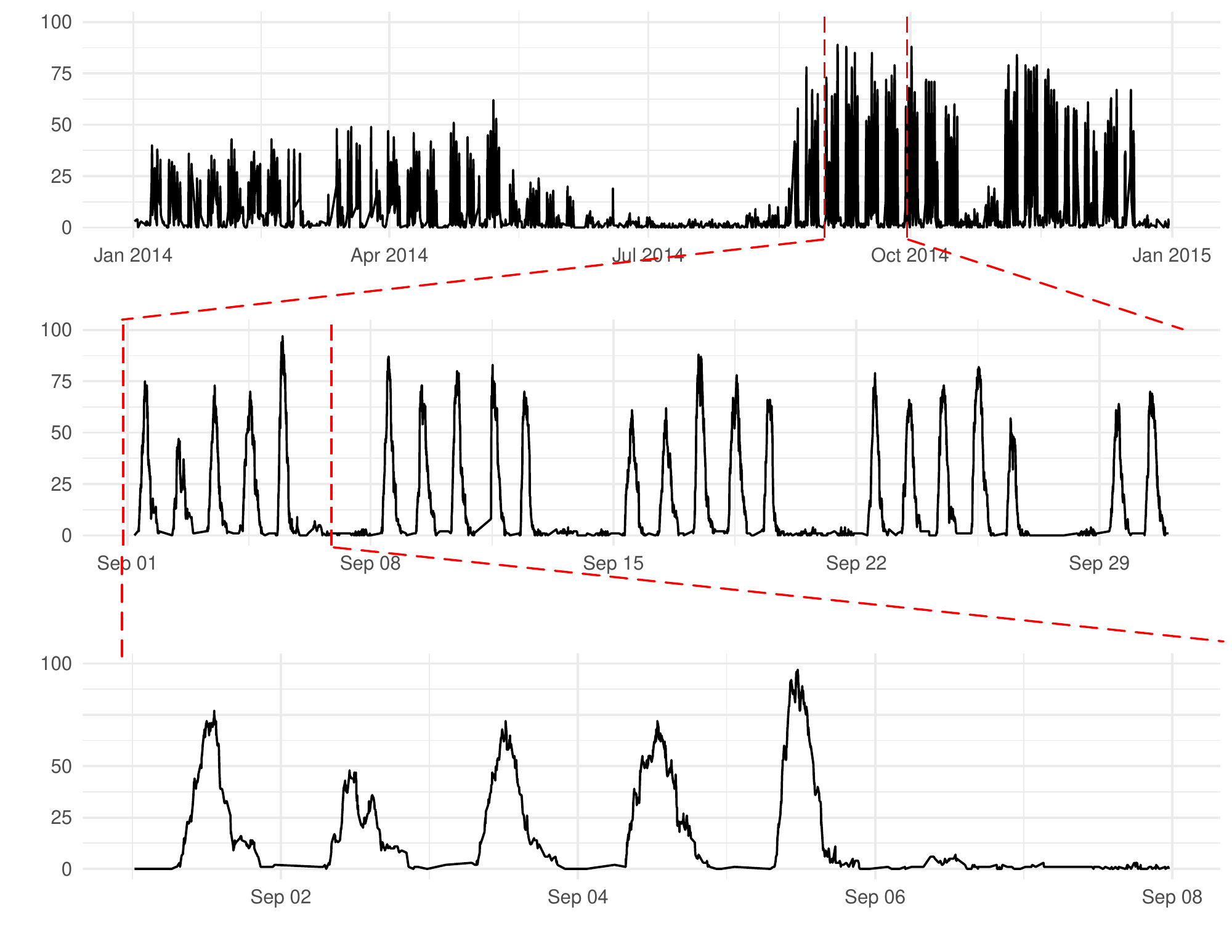} 
  \caption{Time series of the workload of one AP. Topmost graph shows the yearly workload, middle one a month, and the lowermost that of one week.}
  \label{bs_timeseries}
\end{subfigure}
   \caption{Location and workload of each AP. }
\end{figure*}



\hl{The evaluations for the MEC and Fog computing scenarios are conducted as simulations, based on a real-world data set, collected from panOULU public network Wi-Fi APs in the city of Oulu between the years 2007-2015 }\cite{kostakos2013}\hl{. As presented in section }\ref{rel_work}\hl{, related work also present data sets on the same time period. }

In the evaluation, we concentrate on data collected during the full year 2014, as it was observed to have a maximum number of yearly Wi-Fi connections (in total 257,552,871) on 450 active APs. The AP deployment, as shown in Figure \ref{bs_workloads}, consists of both a densely covered city center and sparsely covered suburban areas in the City of Oulu, Finland. 

\hl{Each edge server can handle limited computational load. We assume that each connected user brings a load of one on the edge server. Aggregating a unitary user workload over the hundreds of millions of connections in the recorded data offers a simple yet effective method for estimating the relative workloads between APs, used in the offline edge server placement.}


\hl{Figure }\ref{bs_timeseries}\hl{ illustrates the number of simultaneously connected users in an APs of a local polytechnic as a time series in different scales, i.e., yearly, monthly, and weekly. For example, on the weekly connections panel (bottom), the difference between working days and weekend can be clearly seen. On the yearly panel (top), the concentration of the connections at the start of the semester, after the summer months, is clearly visible. The highest number of simultaneous connections in 2014 for that particular AP reaches ca. 100.}


We define the workload of an AP as the highest number of simultaneous connections to that AP in 2014. This way we emphasize the maximum workload during peak hours. As shown in Figure~\ref{ap_map_hist}, the distribution of such maximum workloads is heavily skewed, with a large number APs with low workloads and small number of APs with exceedingly high workloads.


\hl{PACK admits any pairwise distance metric. However, as with all related work, the topology of the underlying network was not available in the data set. In a heterogeneous data set such as the one used here, consisting of central and remote areas, the geographical distances between hops vary significantly. Such heterogeneity is typical, for example, in data sets featuring metropolitan areas. Typically, a single hop corresponds to a longer geospatial distance in a suburban area than in a dense city center. This makes geospatial distance a biased estimate of the underlying hop distance. However, the center areas with heavy workloads tend to overshadow the remote areas with less workload, resulting in unequal QoS between them. To favor the suburban areas, we exaggerate the distances there by using the squared Euclidean distance as an estimator of the underlying topology.} 

\subsection{Edge server placement for MEC}
\label{mec_placement}


The MEC scenario is based on Multi-access Edge Computing, with medium scale data centers that are commonly co-located with mobile network base stations \cite{dolui2017,ETSI_SwForMEC_whitepaper}, and more recently APs. Powerful edge servers, as a part of the standardized MEC infrastructure, control the computational and communication resources of the connected network infrastructure components, including base stations, APs, switches and routers, etc., each with a set of connected UEs. The resulting edge computing infrastructure is therefore based on a small set of geographically scattered clusters, with powerful edge servers as cluster heads, each with a large number of connected APs.

The MEC scenario assumes a budget of 20 edge servers for simulation, based on approximation with domain expertise. We center the capacity limits around the even workload (426.5) with a relatively wide window width, resulting in a lower capacity limit 327 and upper limit 526. We compare six different MEC server placement setups: 
\begin{enumerate}[label=(\alph*)]
    \item[\textit{M1}] Both lower and upper capacity constraints are present. 
    \item[\textit{M2}] \hl{No capacity constraints at all, which corresponds to the discrete k-means method, also referred to as the k-medoid method.}
    \item[\textit{M3}]  \hl{Only upper capacity constraint, which was the typical way to balance the workload in the literature}.
    \item[\textit{M4}] Both lower and upper capacity constraints are present with fractional membership for the APs. 
    \item[\textit{M5}] Both lower and upper capacity constraints are present with location preference set for five APs.  
    \item[\textit{M6}] Both lower and upper capacity constraints are present with enhanced reliability for five APs.
\end{enumerate}

\hl{Except with setup \textit{M4}, we default to hard membership, assuming the edge servers will not share the AP workloads.} In \textit{M5} and \textit{M6}, we assume that the APs are selected based on domain expertise and may be, for example, part of critical infrastructure. Location preference is achieved by setting the parameter $\gamma_i = 100$ for the selected APs. Reliability, by way of replicated workload, is achieved by assigning five APs to exactly two servers, i.e. $q_i=2$ for the 5 critical APs.



\hl{Quantitative evaluation metrics (see section }\ref{eval_metrics}\hl{) are collected in Table }\ref{QoS_table_MEC}\hl{. Further, the resulting optimal server placement candidates for each setup in the MEC scenario are shown in Figure }\ref{mec_scenarios} and Figure \ref{mec_priority}\hl{, highlighting the trade-off between latency and workload balance.}



\begin{table*}[ht]
\centering
\footnotesize
\begin{tabular}{llll|lllll|lll}
  \toprule
  \multicolumn{4}{l|}{Setup} & \multicolumn{5}{l|}{Relative latency} & \multicolumn{3}{l}{Workload} \\
  \midrule
& k & frac & cap & Mean & 25\% & 50\% & 75\% & 95\% & S.D. & Min & Max \\ 
  \midrule
  \textit{M1} & 20 & no & up/low & 0.59 & 0.06 & 0.28 \cellcolor{green!40}  & 0.81 & 2.11 & 90 & 327 \cellcolor{green!40}  & 526 \cellcolor{green!40}  \\ 
  \textit{M2} & 20 & no & no & 0.40 \cellcolor{green!40}  & 0.02 \cellcolor{green!40}  & 0.29 & 0.48 \cellcolor{green!40}  & 1.33 \cellcolor{green!40} & 336 & 55 & 1013 \\ 
  \textit{M3} & 20 & no & up & 0.55 & 0.10 & 0.38 & 0.71 & 1.61 & 162 & 57 & 526 \cellcolor{green!40}  \\  
  \textit{M4} & 20 & yes & up/low & 0.59 & 0.04 & \cellcolor{green!40} 0.28 & 0.70 & 2.11 & 89  \cellcolor{green!40} & 327 \cellcolor{green!40} & 526 \cellcolor{green!40} \\ 
  \textit{M5} & 20 & no & up/low & 0.59 & 0.10 & \cellcolor{green!40} 0.28 & 0.73 & 2.11 & 89  \cellcolor{green!40} & 327 \cellcolor{green!40} & 526  \cellcolor{green!40} \\ 
  
  \textit{M6} & 20 & no & up/low & 0.63 & 0.10 & 0.30 & 0.85 & 2.19 & 95.2   & 327 \cellcolor{green!40} & 526 \cellcolor{green!40}  \\ 
  \midrule
\bottomrule
\end{tabular}
\caption{Evaluation results of the six MEC setups, comparing relative latency between edge servers and APs, and workload balance on the edge servers. k = number of servers. frac = fractional membership. \\ cap = capacity constraints (up = upper constraint, low = lower constraint). \\  Mean = the average AP distance, weighted by the workload. $\alpha \%$= the maximum distance between AP and edge server for $\alpha\%$ of the workload. Workload balance is measured with the standard deviation of the server workloads (S.D.)  and the minimum (Min) and the maximum (Max) obtained server workload. See Section~\ref{eval_metrics} for further information. The green background indicates the best performance in terms of the different quality measures.}
\label{QoS_table_MEC}
\end{table*}

\hl{Setup \textit{M1} has low latencies for the 25\% and 50\% quantiles, with the median latency ($\alpha = 50\%$) as the lowest of all setups. Further, as both the upper and the lower capacity constraints are applied in the setup, \textit{M1} has excellent workload balance. However, the upper quantile latencies are clearly higher than in the other scenarios, resulting in the highest mean latency. Due to the uneven distribution between workload and AP densities, this combination of good workload balance and high worst case latency is further evident in Figure }\ref{mec_scenarios}\hl{, where the high worst-case latencies manifest as widely differing spatial areas of the edge server clusters: the wider the coverage of an edge server, the worse the latencies to that server from the APs at the borders. Nevertheless, the result shows that PACK does not under- or over-provision server capacity, regardless of the resulting geospatial coverage.}

\hl{Conversely, setup \textit{M2}, with no capacity constraints, has excellent latency with spatially equally sized clusters, but highly varying server workloads with a very high standard deviation. In such a setup, edge servers need to be heterogeneous, with varying capacity, to handle the resulting very low and very high workloads.}

\hl{Setup \textit{M3}, without a lower capacity constraint, presents a middle ground between \textit{M1} and \textit{M3}. Setup \textit{M3} has four servers with workload under 327, the lower limit of scenario \textit{M1}. As a result, \textit{M3} has lower latency than \textit{M1} for the large quantiles. On the other hand, due to the upper capacity constraint, \textit{M3} has better workload balance than \textit{M2}. Accordingly, the resulting clusters vary in spatial size less than those of \textit{M1}, but more than in \textit{M2} (Figure }\ref{mec_scenarios}\hl{).}

\hl{Setup \textit{M4} introduces fractional membership together with both upper and lower capacity constraints. Fractional membership allows the workload of APs to be shared among a number of edge servers. The resulting placement candidate is nearly identical to \textit{M1} in terms of workload balance and latency, with just a slight improvement in the best case QoS, that is, the 25\% quantile of relative latency. This results from the increased leeway in the placement and allocation provided by the fractional membership. Further, Figure }\ref{mec_priority}\hl{ shows some spatial overlap of the clusters due to the division of workload on some APs.}



\begin{table*}[ht]
\centering
\footnotesize
\begin{tabular}{l|lllll}
  \toprule
  Setup & \multicolumn{5}{l}{Relative latency} \\
  \midrule
  & \multicolumn{5}{c}{AP id} \\
   & 800 & 588 & 712 & 974 & 693 \\
   \cmidrule{2-6}
 \textit{M1} & 40.4 & 162.6 & 93.4 & 122.0 & 29.2 \\
 \textit{M5} & 40.4 & 82.9 & 76.3 & 37.7 & 29.2 \\
\midrule
\bottomrule
\end{tabular}
\caption{Relative latency from the five preferred APs to the closest server in scenarios \textit{M1} and \textit{M5}.}
\label{m1vsm5_table}
\end{table*}

\hl{Setup \textit{M5} has five pre-selected, preferred locations for edge servers. As shown in Table }\ref{m1vsm5_table}\hl{, \textit{M5} ends up with three server locations closer to the preferred locations than in scenario \textit{M1}. As a result, the preferred server locations enjoy a much lower latency, at the cost of a slightly worse best-case QoS, that is, the 25\% quantile of relative latency.}

\hl{Setup \textit{M6} replicates the workload of five pre-selected APs on two different edge servers, thus providing those APs with enhanced  reliability. As a result, server clusters are forced to overlap for those APs, and the geospatial areas of those clusters are elongated to cover the replicated APs (see Figure \ref{mec_priority}). While latencies increase slightly throughout, and the workload balance is marginally worsened, enhancing reliability by replication of a few critical APs does not significantly affect QoS across the deployment.}

Illustrating the workloads across the deployment, Figure \ref{mec_workloads} shows the optimal edge server locations resulting from setups (\textit{M1--M4}), and the individual workloads of servers, some of which may be overlapping in the map illustration. Large workloads appear in densely populated areas, e.g. the city center and the university campus area to the north. The sparsely populated areas have significantly less workload. For scenario \textit{M2}, the resulting powerful servers are clearly seen. \hl{Furthermore, a tighter capacity constraints setup is presented in }\ref{Appendix}. \hl{The setup ensures a well balanced workload among servers with standard deviation being the smallest among setups, however the overall latency is worse when compared to \textit{M1}.}

\subsection{Large-scale Fog deployment}
\label{fog_placement}


\hl{The Fog scenario is based on the Fog computing architecture, which provides a virtualized, hierarchically organized, distributed computing platform, based on small-scale Fog nodes at or near APs, such as common household appliances, smartphones, laptops, and set-top boxes, as well as network infrastructure components on the higher echelons of the hierarchy }\cite{dolui2017}\hl{. The Fog nodes close to UEs are expected to have smaller capacity, increasing towards the cloud in the infrastructure, forming a "cloud-thing continuum" }\cite{Yousefpour2019}\hl{. Fog thus offers low latency for edge applications in largely isolated deployments, e.g., a business facility. The resulting edge platform infrastructure is based on numerous small-sized clusters with lower capacity servers. Further, where high capacity is needed and increased deployment and management costs are acceptable, a cloudlet could be built with an interconnected set of Fog nodes.}

The Fog scenario considers the placement of 150 edge servers, i.e. Fog nodes, with the following capacity constraints: lower limit at 10, and upper limit at 80. The limits are determined with domain expertise: ca. 10 UEs are assumed to be connected to an AP in a typical household, and ca. 80 UEs are assumed to be connected to a more powerful Fog node, covering a building or in an office. Analysis of the original data set indicated that 17 APs had workloads larger than the upper limit, which necessitates the use of fractional membership. 

We compare three different Fog server placement setups:
\begin{enumerate}[label=(\alph*)]
    \item[\textit{F1}] \hl{Both lower and upper capacity constraints are present. Fractional membership.}
    \item[\textit{F2}] \hl{No capacity constraints, fractional membership. This corresponds to the discrete k-mean method with fractional membership.} 
    \item[\textit{F3}] \hl{Only the upper capacity constraint is present, fractional membership. This was the typical way to balance the workload in the literature.}
\end{enumerate}

\hl{Quantitative evaluation metrics (see Section }\ref{eval_metrics}\hl{) can be found in Table }\ref{tab:QoS_table_Fog}\hl{. Figure }\ref{Fog_clustering}\hl{ further illustrates the resulting optimal edge server locations for these scenarios, while Figure }\ref{Fog_hist}\hl{ shows the distributions of workload on the edge servers as histograms for all Fog setups.}

\begin{table*}[ht]
\centering
\footnotesize
\begin{tabular}{llll|lllll|lll}
  \toprule
  \multicolumn{4}{l|}{Setup} & \multicolumn{5}{l|}{Relative latency} & \multicolumn{3}{l}{Workload} \\
  \midrule
& k & frac & cap & Mean & 25\% & 50\% & 75\% & 95\% & S.D. & Min & Max \\ 
  \midrule
  \textit{F1} & 150 & yes & up/low & 0.05 & 0 \cellcolor{green!40} & 0 \cellcolor{green!40} & 0.02 & 0.26 & 28  \cellcolor{green!40} & 10 \cellcolor{green!40} & 80 \cellcolor{green!40} \\ 
  \textit{F2} & 150 & yes & no & 0.02 \cellcolor{green!40}& 0 \cellcolor{green!40} & 0 \cellcolor{green!40} & 0 \cellcolor{green!40} & 0.09 \cellcolor{green!40} & 77 & 1 & 612 \\ 
  \textit{F3} & 150 & yes & up & 0.05 & 0 \cellcolor{green!40} & 0 \cellcolor{green!40} & 0.02 & 0.28 & 29 & 1 & 80 \cellcolor{green!40} \\ 
  \midrule
\bottomrule
\end{tabular}
\caption{Evaluation results of the three Fog setups, comparing relative latency between edge servers and APs, and workload balance on the edge servers. k = number of servers. frac = fractional membership. \\ cap = capacity constraints (up = upper constraint, low = lower constraint). \\  Mean = the average AP distance, weighted by the workload. $\alpha \%$= the maximum distance between AP and edge server for $\alpha\%$ of the workload. Workload balance is measured with the standard deviation of the server workloads (S.D.)  and the minimum (Min) and the maximum (Max) obtained server workload. See Section~\ref{eval_metrics} for further information. The green background indicates the best performance in terms of the different quality measures.}
\label{tab:QoS_table_Fog}
\end{table*}

\hl{Setup }\textit{F1}\hl{, with both upper and lower capacity constraints, prefers workload balance at the cost of latency. About half of the servers (72) have their workloads at the maximum of 80. The distribution of workload (Figure }\ref{Fog_hist}\hl{) is distinctly double peaked, with one peak close to the lower limit and the other close to the upper limit. The reason for the double peaks, is that the workloads tend to be high in the city center, i.e. servers with maximum capacity are needed, whereas in the suburban areas the workloads are close to the minimum (Figure }\ref{Fog_clustering}\hl{). \textit{F1} has a heterogeneous deployment across the area, with low capacity servers fulfilling the latency constraint for the sparse areas, and a dense deployment of high capacity servers in the city center (Figure }\ref{Fog_clustering}\hl{.}

\hl{Setup }\textit{F2}\hl{, conversely, prefers latency at the cost of workload balance. Indeed, with no capacity constraints, an AP is always assigned to the closest server, resulting in the lowest latencies for all the quantiles, as well as the mean. However, the setup results in a high variation in server workloads, with 29 servers, or 80\% of the total deployment, exceeding the upper constraint of 80 set for other setups. The highest workload is 612, with the distribution heavily skewed to the right (Figure }\ref{Fog_hist}\hl{). The servers with the heavy loads are concentrated in the city centre and the university campus area to the north, with the low-capacity servers populating the sparse areas (Figure }\ref{Fog_clustering}\hl{).}

\hl{Due to the low value of the lower capacity constraint, setup \textit{F3} results in a very similar placement candidate as \textit{F1}. Figure }\ref{split_workload}\hl{ suggests a tendency towards more shared workload in the dense regions and less in the sparse regions, when compared with \textit{F1}.}

%% file: 5_discussion.tex
\section{Discussion}
\label{discussion}


\hl{The MEC and Fog scenarios and their various setups illustrate the effect of the latency--workload trade-off, and how to control it, as well the effects of the workload sharing with fractional membership, workload replication, and location preference. While the evaluation used a single data set, it was based on a real-world Wi-Fi setup, and the AP and workload densities in the dataset varied immensely. Arguably, the heterogeneity of the data set guarantees the generalizability of the results.}

\hl{Indeed, the trade-off between workload and latency can clearly be observed in the results. Using both upper and lower capacity constraints results in a well-balanced load across a heterogeneous data set, but with variance in the resulting latencies, manifesting as highly variable spatial coverage of the AP clusters. Conversely, no capacity constraints results in excellent latency, with AP clusters of equal spatial coverage, but high variance in the resulting workload on the edge servers.} 

\hl{For MEC, fractional membership does not appear to have a strong effect on the resulting deployment. Regarding latency as a measure for QoS, the placement without capacity constraints produces the best results in this respect as it ignores load balancing. If only the upper capacity constraint is set, latency becomes worse for the lower quantiles, while workload balance is much improved. Adding also the lower capacity constraint, the workload balance is somewhat improved, while latency slightly worsens. Alternatively, allowing diversity in the capacity of the edge servers may help in reducing the unused capacity typically caused by applying only the upper capacity limit. Further, when a small number of critical servers are assigned a higher location preference or provided high reliability, the effect is minimal on overall latency and workload balance.} 

\hl{The Fog scenario results give three major insights on the hardware configurations and their capacities needed for the Fog nodes, helping to determine the required budget for the Fog deployment. First, omitting capacity constraints in the Fog scenario leads to unrealistically high workloads on some Fog nodes. Capacity constraints are thus necessary at the lowest layer in the Fog hierarchy. Second, for dense areas, with many users and APs, a deep hierarchy of nodes should be deployed, with increasingly capable nodes on the higher levels. On the other hand, in sparser populated areas, it is sufficient to have a flat hierarchy, with a high number of less capable nodes. Third, for the constrained setups in the Fog scenario, depending on the deployment budget, two levels of Fog node hierarchy would be sufficient, as two peaks were seen in the results: one level of light servers for the peak in the lower workloads, and another, more capable one for the peak on the higher workloads. This simplifies the deployed Fog node configurations.}

The possibility to share the workloads between edge servers makes the PACK available for wide variety of deployment scenarios. In Fog, i.e. dense deployments of low capacity servers, the benefit of this relaxation is emphasized when the workload of an existing AP is greater than the capacity limit of the associated edge server(s). This way, partial workload can be handled in premises, while the rest is horizontally shared. 

In the MEC scenario, shared workload did not have such an effect. This is due to the distribution of the workloads of APs: the average workload in sparse regions is relatively small compared to the predefined capacity range, which reduces the need for sharing. Indeed, fractional membership is more useful when tighter capacity limits are applied, and the average workload in sparse areas is larger. In such cases, finding an optimized solution with hard membership takes more computational time and, in some cases, a solution may not even be found. 



Adding to the practical deployment parameters of PACK, location preference is an option in real-world deployments. Certain facilities may require a server, e.g. a research facility, that may or may not be shared with the users of the mobile network. Such prefixed locations should be taken into account in finding the optimal solutions, since the effects are seen in placing of the other servers, and on how the workload is balanced. Such servers could be left out of the initial deployment altogether, and later used as a reserve capacity. 

Further, location preference can be considered from the critical network infrastructure point-of-view, where selected APs are given more weight in the placement algorithm. This approach demonstrates an easy but effective solution for addressing location preference in placement algorithms. Furthermore, it enables one to easily allow several preference classes with different weights.

Similarly, reliability is a considerable pragmatic challenge in edge computing in general. As demonstrated, with PACK and in the previous work, a way to address reliability is to add duplicate capacity that becomes operational in case of server failures. Naturally, such an approach further increases initial deployment, maintenance and management costs. In placement algorithms, a feasible option, as demonstrated, is to define a set of servers as a part of the critical infrastructure, and provide duplicate capacity only for them.


Overall, edge server placement is a problem of increasing importance in future distributed IoT systems. We believe that the PACK algorithm, with extensive set of deployment parameters, based on significantly large data set and detailed statistical evaluation in a multiple scenarios, simultaneously and comparably, is a noteworthy addition to the knowledge in the field. 



%% file: 6_conclusion.tex
\section{Conclusion}
\label{conclusion}

Solving the edge server placement problem provides the foundation for further study of other edge resource provisioning problems, such as online load balancing. In this paper, we first conducted a survey of existing physical edge server placement algorithms and identified an extensive set of parameters utilized in placement. Based on the findings, we presented the PACK algorithm that considers the parameter set in optimal placement for edge servers with load balancing under capacity constraints, while minimizing latencies between the servers and their associated APs. 

We evaluated the PACK algorithm, and studied the effects of different placement parameters, with simulations in two distinct edge deployment scenarios, corresponding to MEC and Fog computing. The simulations were based on a large real-world data set, collected in a city-wide Wi-Fi network between the years 2007-2015.

The evaluations provided numerical statistics on QoS and QoE for analyzing the performance of PACK in each scenario. 
The latency--workload balance was illustrated with extensive examples, and methods for controlling the trade-off were detailed. The influence of workload sharing, workload replication and location preference on the placement were also discussed in detail.

Further, the evaluations grand broad insights on how well the algorithm performs with different architectural design options and on planning of deployment setups with a set of constraints. Edge computing systems are developing towards large numbers of distributed, small capacity computational units, deployed into individual sites, as in Fog computing. Moreover, in evaluation, a number of pragmatic challenges for mobile network operators were taken into account, providing insights on how well the algorithm performs under real-world constraints. 



We argue that the flexibility of the PACK algorithm, combined with the easy to use open-source R based software package, provides an analytical toolbox to study different edge server placement schemes, before making the actual deployment decisions. 
Such a tool box is a key component towards the future vision of edge-native artificial intelligent solutions \cite{Loven2019edgeai1}, providing the theoretical background to find optimal solutions to the placement problem, with real-world edge applications and fine-grained data collected on their performance.

%% file: 7_acknowledgments.tex
\section*{Acknowledgments}

This research is supported by Academy of Finland 6Genesis Flagship (grant 318927), the Infotech Oulu research institute, the Future Makers program of the Jane and Aatos Erkko Foundation and the Technology Industries of Finland Centennial Foundation, by Academy of Finland Profi 5 funding for mathematics and AI: data insight for high-dimensional dynamics (grant 326291), and the personal grant for Lauri Lovén on Edge-native AI research by the Tauno Tönning foundation.


%% file: appendix_tight.tex
\appendix
\section{Scenario with tight capacity limits}
\label{Appendix}

\begin{figure}
    \centering
    \includegraphics[width=1\linewidth]{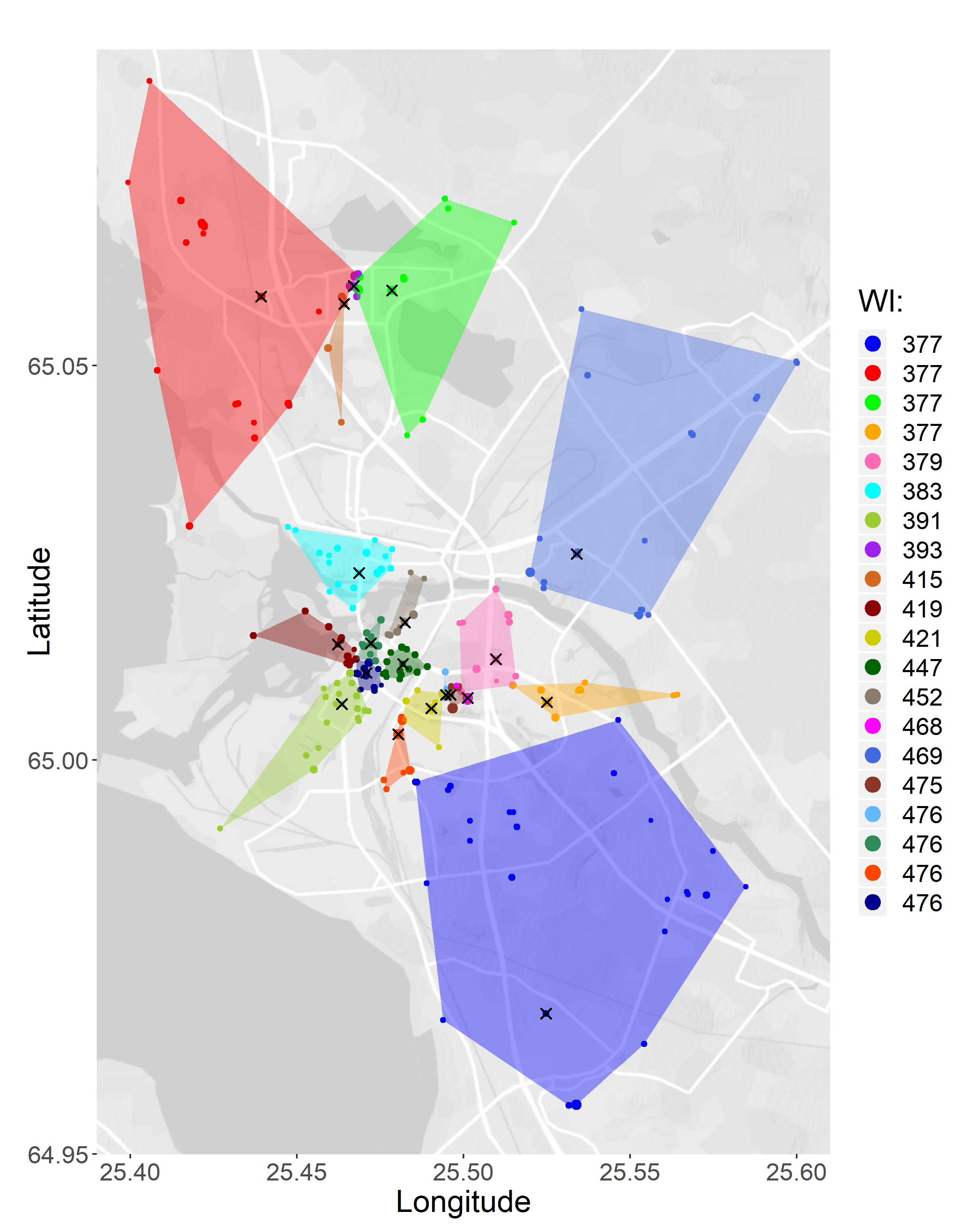}
    \caption{Placement of 20 MEC servers with tight capacity limits. Crosses denote the server locations. Wl: workloads of the servers. The color shades visualize the server regions.}
    \label{tight_cap}
\end{figure}

In this section we consider a MEC placement scenario where both lower and upper capacity constraints are set. We set a relatively tight width for the capacity, resulting in a lower capacity limit of 377 and upper limit of 476. The quantitative measures are summarized in Table \ref{QoS_table_app} and the resulting edge server placement of the scenario is presented in Figure \ref{tight_cap}.


\begin{table*}[ht]
\centering
\footnotesize
\begin{tabular}{llll|lllll|lll}
 \multicolumn{4}{l|}{Setup} & \multicolumn{5}{l|}{Relative latency} & \multicolumn{3}{l}{Workload} \\
  \toprule
& k & frac & cap & Mean & 25\% & 50\% & 75\% & 95\% & S.D. & Min & Max \\ 
  \midrule
  \textit{M1} & 20 & no & up/low & 0.59 & 0.06 & 0.28 \cellcolor{green!40}  & 0.81 & 2.11 & 90 & 327   & 526  \\ 
  \textit{M2} & 20 & no & no & 0.40 \cellcolor{green!40}  & 0.02 \cellcolor{green!40}  & 0.29 & 0.48 \cellcolor{green!40}  & 1.33 \cellcolor{green!40} & 336 & 55 & 1013 \\ 
  \textit{M3} & 20 & no & up & 0.55 & 0.10 & 0.38 & 0.71 & 1.61 & 162 & 57 & 526   \\  
  \textit{M4} & 20 & yes & up/low & 0.59 & 0.04 & 0.28 & 0.70 & 2.11 & 89   & 327  & 526 \\ 
  
  \textit{M5} & 20 & no & up/low & 0.59 & 0.10 & 0.28 & 0.73 & 2.11 & 89  & 327  & 526  \\ 
  
  \textit{M6} & 20 & no & up/low & 0.63 & 0.10 & 0.30 & 0.85 & 2.19 & 95.2   & 327  & 526   \\ 
  
  Tight & 20 & no & up/low & 0.63 & 0.04 & 0.30 & 0.75 & 2.25 & 41.9 \cellcolor{green!40} & 377 \cellcolor{green!40} & 476 \cellcolor{green!40}\\ 
  
    \midrule
\bottomrule
\end{tabular}
\caption{Comparing the distance and balance of the nine different deployment scenarios (see Section ~\ref{mec_placement}). k = number of servers. frac = fractional membership. \\ cap = capacity constraints (up = upper constraint, low = lower constraint). \\  Mean = the average AP distance weighted by the workload. $\alpha \%$= the distance within which $\alpha\%$ of the workload is from the associated server. See Section~\ref{eval_metrics} for further information. The workload balance is measured with the standard deviation of the server workloads (S.D.)  and the minimum (Min) and the maximum (Max) obtained server workload. The green background indicates the best performance in terms of the different quality measures in the MEC scenarios \textit{M1 - M6} and the tight capacity limits scenario.}
\label{QoS_table_app}
\end{table*}

In terms of latency, the tight capacity limits scenario performs similarly to the scenario \textit{M6} as can be seen in the mean value in Table \ref{QoS_table_app}. Furthermore, the scenario has the worst worst-case latency quantile (95\%). This is confirmed in the Figure \ref{tight_cap}, where the server regions are large in the sparse regions. However, it is clear that the tight capacity limits ensure a well balanced workload with approximately 0.5 times smaller standard deviation than in the other capacitated scenarios in Table \ref{QoS_table_app}. 

\begin{figure*}[ht!]
\centering
\captionsetup[subfigure]{labelformat=empty}
\begin{subfigure}[t]{.49\textwidth}
    \centering
    \includegraphics[width=1\linewidth]{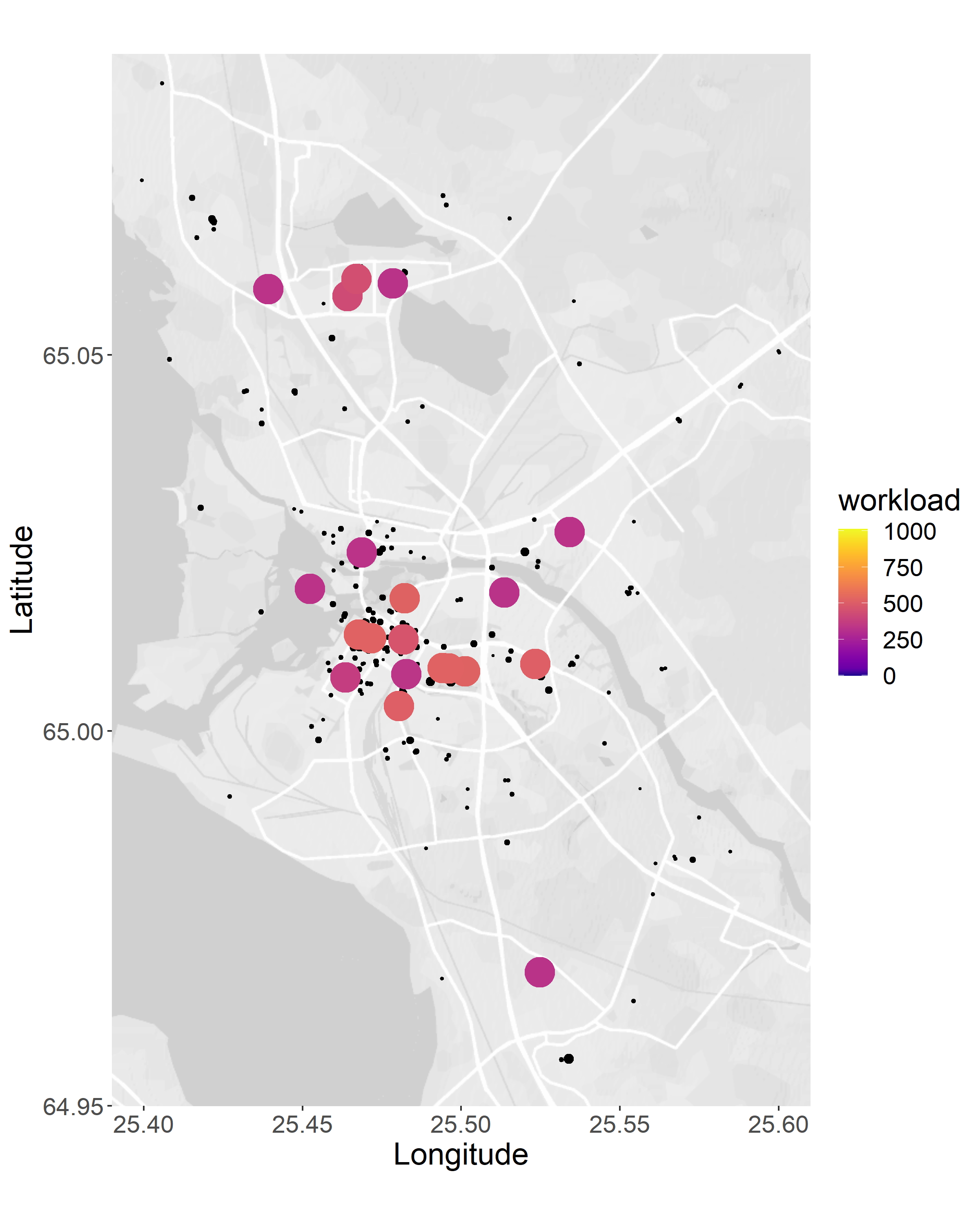}
    \caption{\textbf{\textit{M1} Upper and lower capacity constraints.}}
\end{subfigure}
\begin{subfigure}[t]{.49\textwidth}
  \centering
    \centering
    \includegraphics[width=1\linewidth]{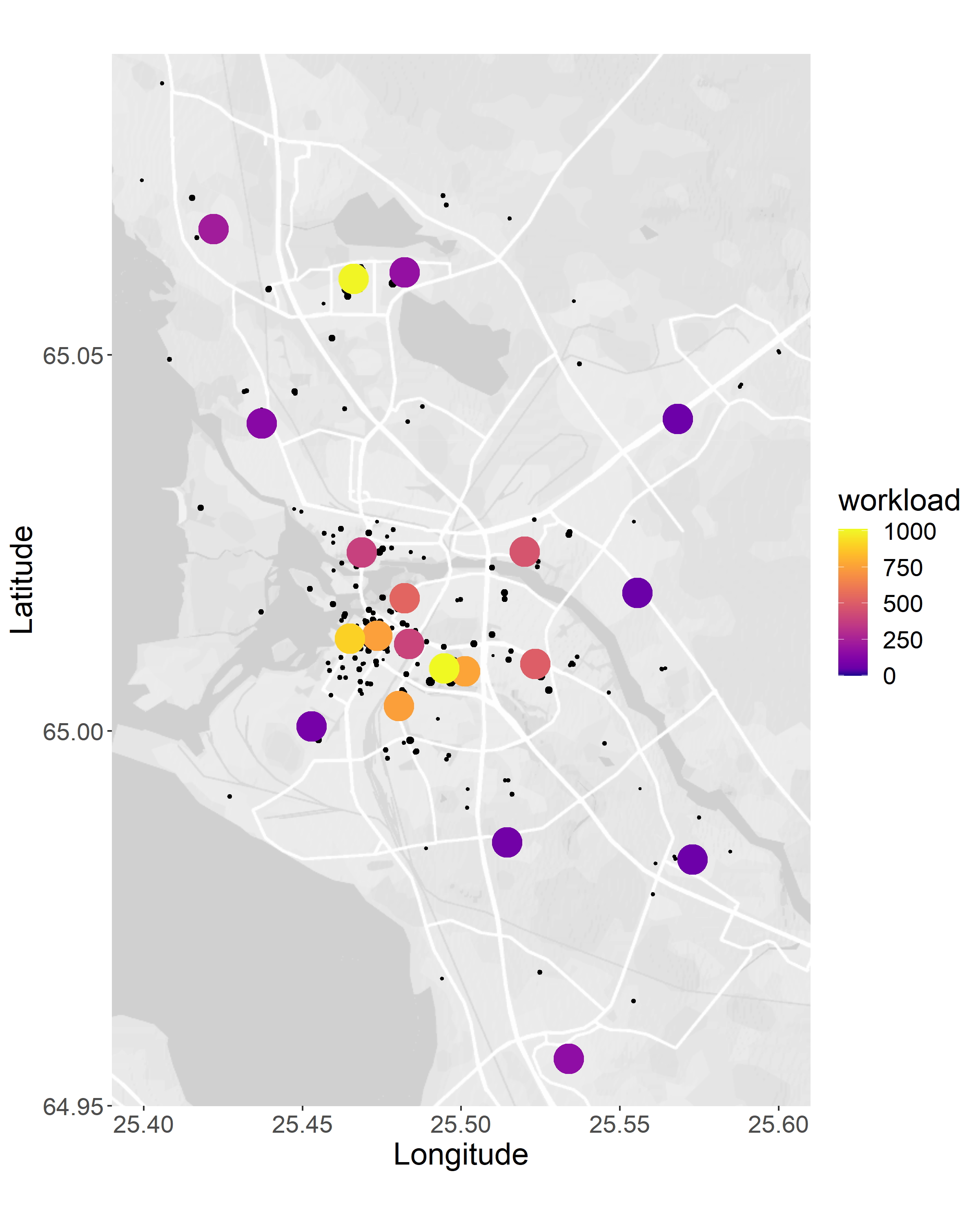}
    \caption{\textbf{\textit{M2} No capacity constraints.}}
\end{subfigure}
\begin{subfigure}[t]{.49\textwidth}
  \centering
    \centering
    \includegraphics[width=1\linewidth]{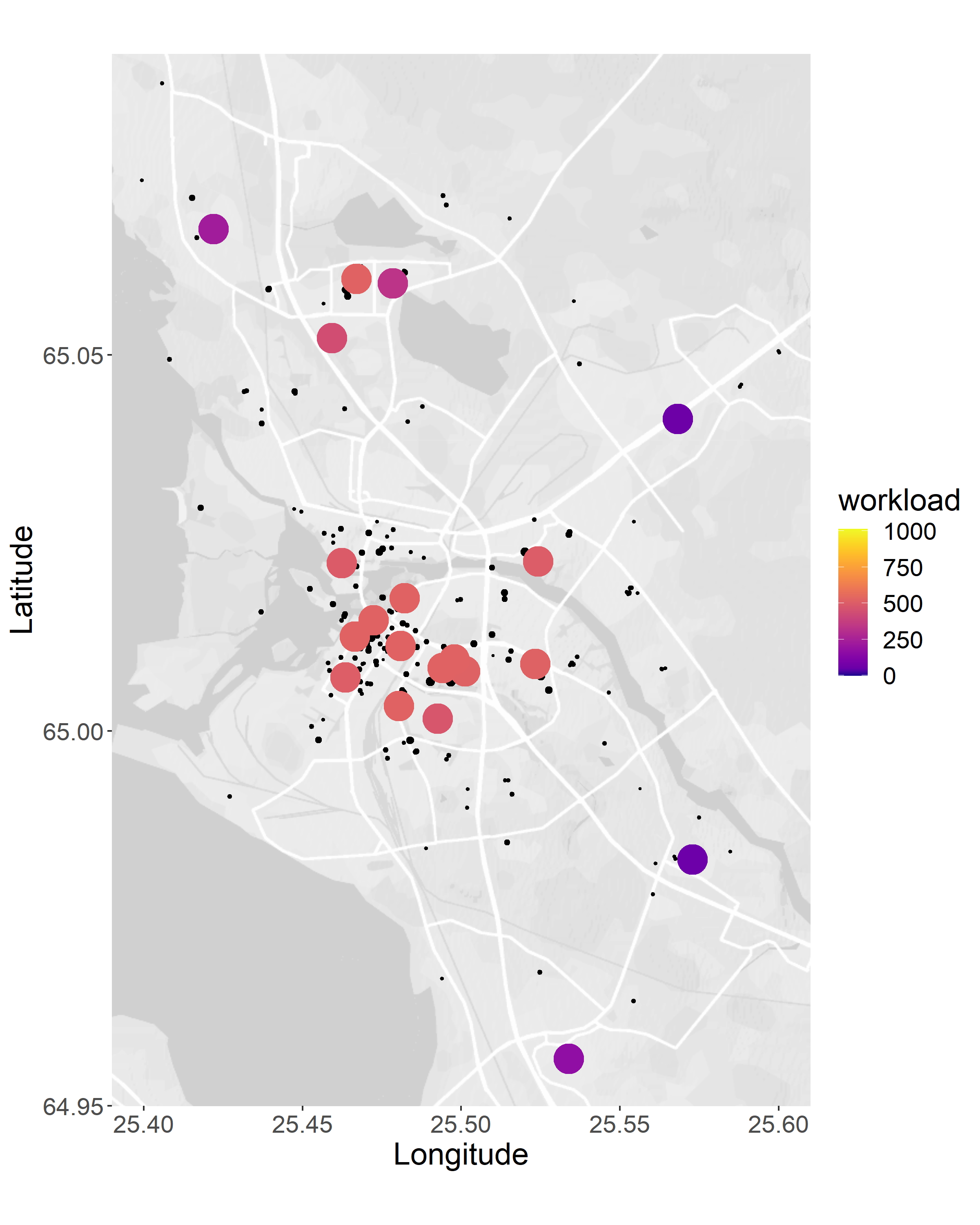}
    \caption{\textbf{\textit{M3} Only upper capacity constraint.}}
\end{subfigure}
\begin{subfigure}[t]{.49\textwidth}
  \centering
    \centering
    \includegraphics[width=1\linewidth]{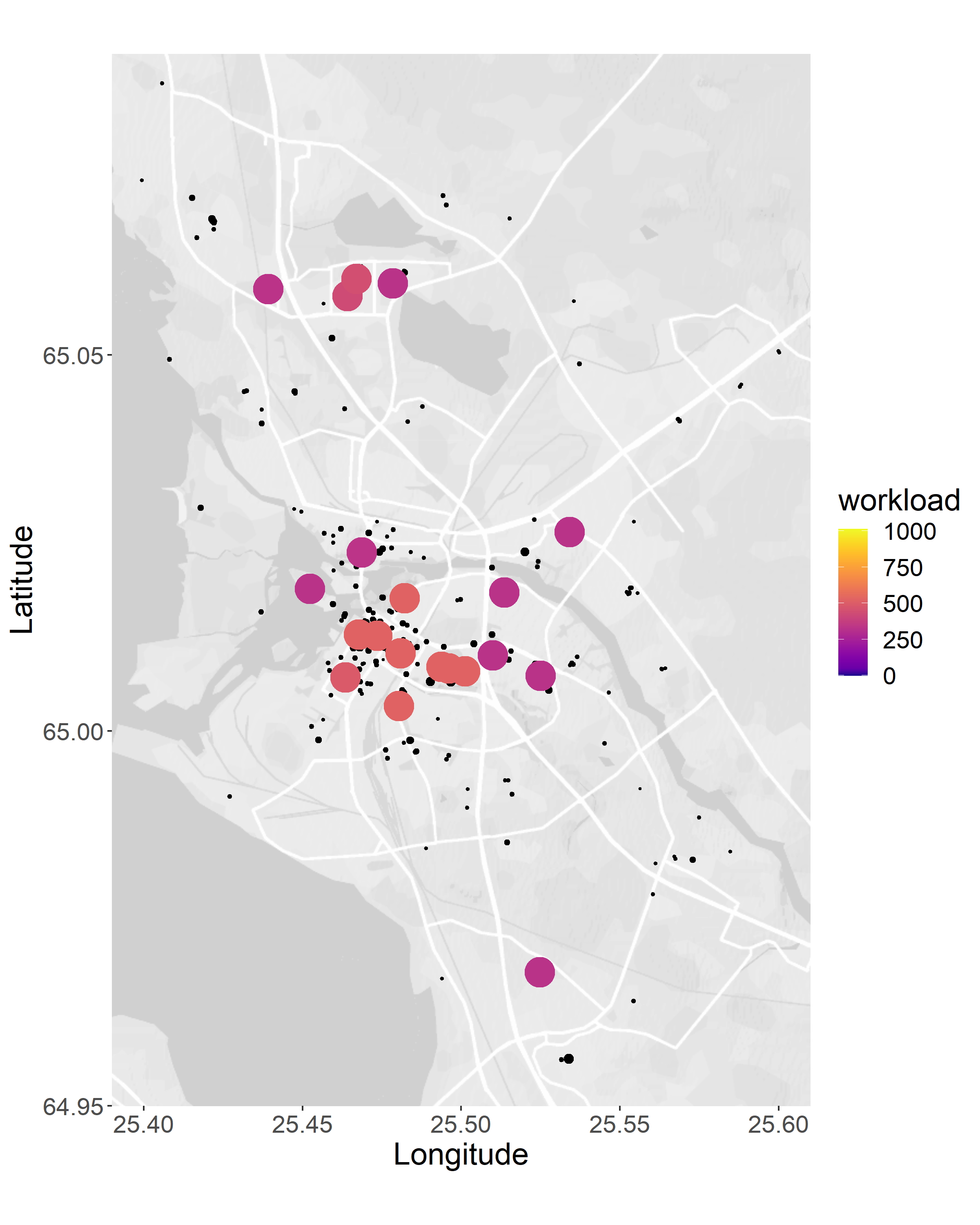}
    \caption{\textbf{\textit{M4} Fractional membership.}}
\end{subfigure}
\caption{Assigned workload of each MEC server. Black dots indicate the APs and colored dots represent the servers, the  shade of color indicating the workload (Wl).}
    \label{mec_workloads}
\end{figure*}